\documentclass{aa}  
\usepackage{graphicx}
\usepackage{txfonts}
\usepackage{graphicx}
\usepackage{amsmath}
\usepackage{hyperref}
\usepackage{natbib}
\usepackage{url}
\usepackage{times}
\def\spose#1{\hbox to 0pt{#1\hss}}
\newcommand\lsim{\mathrel{\spose{\lower 3pt\hbox{$\mathchar"218$}}
     \raise 2.0pt\hbox{$\mathchar"13C$}}}
\newcommand\gsim{\mathrel{\spose{\lower 3pt\hbox{$\mathchar"218$}}
     \raise 2.0pt\hbox{$\mathchar"13E$}}}

\makeatletter
\renewcommand*\aa@pageof{, page \thepage{} of \pageref*{LastPage}}
\makeatother

\begin{document} 

\title{How to constrain mass and spin  of supermassive black holes through their disk emission}
	\titlerunning{SMBH mass and spin constraints}
\author{Samuele Campitiello\inst{1}\thanks{\email{sam.campitiello@gmail.com}}
\and Gabriele Ghisellini\inst{2}
\and Tullia Sbarrato\inst{1}
\and Giorgio Calderone\inst{3}}

\institute{Dipartimento di Fisica "G. Occhialini", Università di Milano - Bicocca, Piazza della Scienza 3, I-20126 Milano, Italy
   \and INAF - Osservatorio Astronomico di Brera, via E. Bianchi 46, I-23807, Merate, Italy
\and INAF - Osservatorio Astronomico di Trieste, via Tiepolo 11, I-34131, Trieste, Italy
            }

\date{Received ; accepted }

\abstract{We investigate the global properties of the radiation emitted by the accretion disk around Kerr black holes. Using the Kerr blackbody (KERRBB) numerical model, we build an analytic approximation of the disk emission features focusing on the pattern of the produced radiation as a function of the black hole spin, mass, accretion rate and viewing angle. The assumption of having a geometrically thin disk limits our analysis to systems emitting below $\sim 0.3$ of the Eddington luminosity. We apply this analytical model to four blazars (whose jets are pointing at us) at high redshift that show clear signatures of disk emission. For them, we derive the black hole masses as a function of spin. If these jetted sources are powered by the black hole rotation, they must have high spin values, further constraining their masses.}

\keywords{galaxies: active -- (galaxies:) quasars: general -- black hole physics -- accretion, accretion disks}
\maketitle


\section{Introduction} \label{sec:intro}

A growing number of supermassive black holes (SMBHs) with masses $M \sim 10^{8}-10^{10} M_{\odot}$ have been observed at high redshifts, but the issue of how they grew so rapidly is still open and under debate. Two possible scenarios can be used in this context: merging between two or multiple black holes, and accretion of material onto the black hole at a high rate (e.g., \citealt{HaiLoe}; \citealt{Yoo}; \citealt{VolRees}; \citealt{LiYetal}; \citealt{Pelup}; \citealt{TanHai}). In the latter scenario, accretion affects both the black hole mass and its angular momentum (\citealt{Bardeen}): an initially nonrotating black hole reaches its maximum spin when it has roughly doubled its mass through coherent accretion (\citealt{Thorn74}). This suggests that high-redshift SMBHs should spin rapidly due to their large mass assembled in less than a Gyr (i.e., at redshift $\sim$6). The high angular momentum is a crucial ingredient for the production of powerful relativistic jets (\citealt{BlandZna}, \citealt{Tchek}).

Knowing the mass and spin of SMBHs is necessary to better understand their physics and evolution, even in connection with their host galaxies (e.g., \citealt{Magor}, \citealt{Gebhar}, \citealt{Ferra}, \citealt{Marconi}, \citealt{Haring}, \citealt{Gultek}, \citealt{Beifiori}, \citealt{Kormen}, \citealt{Mcconn}, \citealt{Reines}). A common evolution is suggested, requiring some form of feedback (see the review by \citealt{Fabian} and references therein). However, the widely applicable black hole estimation methods still suffer from large uncertainties.

As for the ladder of distances, there is a similar ladder concerning black hole masses. There are primary methods based on masers within the accretion disk (see, e.g., \citealt{Greene}, \citealt{Kuo} and the review by \citealt{Tarchi}) or indirect methods based on broad emission line widths, gas kinematic tracing and reverberation mapping (e.g., \citealt{Blandfo}, \citealt{Peterson}, \citealt{PetersonFerr}, \citealt{Onkenetal}; \citealt{FerraFord} and references therein; \citealt{VesterPeter}; \citealt{Bentz}) leading to the {\it virial} mass (e.g., \citealt{Vester}, \citealt{MclureJar}, \citealt{GreenHo}, \citealt{VesterOs}, \citealt{Mcgill}, \citealt{Wangetal}, \citealt{Shenetal11}). The latter methods bear large uncertainties ($\ga 0.4 - 0.5$ dex) that are built on the main calibrating correlations used to find the virial mass, and on their systematics (\citealt{PetersonFerr}, \citealt{Collin}, \citealt{Shenetal}, \citealt{Marconietal}, \citealt{Kelly}, \citealt{ShenKell}, \citealt{Park}).

An alternative method is based on the fit of the accretion disk spectrum. In the simplest case, it depends only on the black hole mass $M$ and the accretion rate $\dot M$, related to two observables: the disk total luminosity and its peak frequency (hereafter ``SED fitting method''; see \citealt{Caldero}). This method is based on the standard alpha-disk, geometrically thin, optically thick down to the innermost radius described by \citealt{SS} (hereafter SS), around a nonrotating black hole (BH) and with no relativistic effects included. By fitting the optical/UV SED of AGNs, several authors have determined the black hole mass and the accretion rate (e.g., \citealt{Shields}; \citealt{MalkSarg}; \citealt{Malkan}; \citealt{SunMalk}; \citealt{Zhengetal}; \citealt{Trakhten}; \citealt{GhiTav09}; \citealt{Sbaretal13}; \citealt{Caldero}, \citealt{Capel1}, \citealt{Capel3}). Other authors included relativistic effects to describe the emission from the blackbody-like disk (e.g., \citealt{NovTho}, \citealt{PageTho}, \citealt{Riff}). 

However, the origin of the so-called ``big blue bump'' emission in AGNs is still under debate. As discussed extensively in \citet{Kora} and references therein, simple thermal models do not provide a good description of the accretion disk emission.
As these authors point out, some of the most significant inconsistencies of the standard SS disk are:

$\bullet$ The broadband continuum slopes $\alpha_{\nu}$ (with $F_{\nu} \propto \nu^{\alpha_{\nu}}$) at optical/near-UV wavelengths found in the literature (e.g., \citealt{Neugetal}; \citealt{Vandenetal}; \citealt{Bonningetal}; \citealt{DavWooBla}) are incompatible with the slope $\alpha_{\nu} = 1/3$, expected from the accretion disk model. We note, however, that the 1/3 slope can be seen at frequencies much lower than the peak, namely in near-IR/optical, where other components contribute (see, e.g., \citealt{Kishietal}, \citealt{Caldero}).

$\bullet$ The spectrum from a simple accretion disk does not reproduce the observed power law extending at X--rays and the soft X--ray excess. This is explained by associating the X--ray emission to a hot corona sandwiching the accretion disk itself (see, e.g., \citealt{Pounds}; \citealt{NanPou}; \citealt{FabianMiniu}).

$\bullet$ 
AGNs with different disk luminosities peak, on average, at similar frequencies (\citealt{Sandersetal}; \citealt{WaltFink}, \citealt{DavWooBla}; \citealt{LaoDav}), but it might not be an evidence against the SS model.

$\bullet$ Microlensing observations in AGNs suggest that the accretion disks are larger than the standard $\alpha$--disk model predicts. \citet{RauchBland}, using Scharzschild holes, found that a thermal accretion disk is too large (by a factor $\sim$3) to fit within the microlensing size constraint. \citet{Jarosetal} found instead that accretion disks were consistent with the microlensing constraint, using Kerr holes. \\

The ongoing discussion is being enriched by the recent discoveries of quasars, at redshifts $2<z<6$, showing not only the big blue bump, but also its peak, unaffected by the intervening IGM absorbing material (e.g., \citealt{Shawetal}; \citealt{Aietal} for the case of SDSS J010013.02+280225.8 at $z=6.3$). In several cases, a SS model provides a good fit to these quasars' thermal emission. On the other hand, we are aware that this model is too simple to describe in detail the emission coming from the inner regions of the disk affected by strong relativistic effects. Also, the thickness of the disk must not be neglected: \citet{LaoNet} (hereafter LN89) and \citet{McClint} stated that, in order to be thin ($z_{\rm r}/r < 0.1$, where $z_{\rm r}$ is the disk half-thickness) and to be described by the \citet{NovTho} accretion theory, a disk must have $\ell_{\rm Edd} = L_{\rm d} / L_{\rm Edd} \lsim 0.3$ for $a=0.9982$ ($\ell_{\rm Edd} \lsim 0.25$ for $a<0$). Hence, the results coming from these thin disk models with $\ell_{\rm Edd} > 0.3$ could be unreliable (see also \citealt{Straubetal} and \citealt{Wieletal} for the case of ultra-luminous X--ray sources).\\

A more complete, relativistic emission model is the Kerr blackbody (KERRBB), designed for Galactic binaries and described by \citet{Lietal} (hereafter Li05), who implemented it in the interactive X-ray spectral fitting program XSPEC (see \citealt{ArnaudXPS} and references therein). In this paper we extend this model to SMBHs showing a good match with observational data (see \S \ref{appl}). Since KERRBB describes a thin disk, its results are valid for $L_{\rm d} < 0.3 L_{\rm Edd}$. For larger values, one should use numerical models that account for the disk vertical structure (e.g., slim disks \citealt{Abretal}, \citealt{SadwAbra}; \citealt{Straubetal}; Polish doughnuts, \citealt{Wieletal}).

With our approach, we will introduce simple analytic tools to study the impact of different spin values and inclination angles on observable features of the accretion disks surrounding SMBHs. This is an improvement with respect to early works (e.g., \citealt{Cunnin}, \citealt{Zhangetal2}), mostly in the field of X--ray binary black holes, since these works explored relativistic correction factors for a limited range of black hole spins and disk inclinations. Our work provides analytic approximations for these factors across the full parameter range of $a$ and $\theta$, under the assumption that KERRBB provides a reasonable description of the spectrum emitted by a disk around both stellar and SMBHs.\\

In \S \ref{sec:basic} we describe the basic assumptions of the KERRBB model. In \S \ref{sec:empat} we focus on how to find accurate analytical formulae that can readily describe the emission pattern of the radiation emitted by disks around Kerr holes (including the case of zero spin). Specifically, in \S \ref{scale} we give the prescription for how to scale from stellar black holes to SMBHs. In \S \ref{KERRfamily} we show that an observed spectrum can be reproduced by a family of KERRBB solutions with different black hole masses, accretion rates and spins. In \S \ref{appl} we apply the KERRBB model to four blazars finding how their black hole masses change as a function of their spins. In this work, we adopt a flat cosmology with $H_0=68$ km s$^{-1}$ Mpc$^{-1}$ and $\Omega_{\rm M}=0.3$, as found by Planck Collaboration XIII (2015).


\section{Assumptions on the KERRBB model} \label{sec:basic}

For a classical SS model, the observed bolometric disk luminosity $L^{\rm obs}_{\rm d}$ scales with the viewing angle $\theta$ as
	\begin{equation} \label{SSobslum}
		L^{\rm obs}_{\rm d} (\theta) = 2 \cos \theta\ L_{\rm d} = 2 \cos \theta\ \eta \dot{M} c^2,
	\end{equation}

\noindent where $L_{\rm d}$ is the total disk luminosity and $\eta =R_{\rm g} / 2 R_{\rm in}\sim 0.083$ is the Newtonian radiative efficiency ($R_{\rm g}$ is the gravitational radius). As shown in \citet{Caldero}, the peak frequency $\nu_{\rm p}$ and luminosity $\nu_{\rm p} L_{\nu_{\rm p}}$ of the observed disk spectrum scale with the black hole mass and accretion rate as
	\begin{equation} \label{eq:nupeak}
		\frac{\nu_{\rm p}}{\text{[Hz]}} = \mathcal{A} \left[ \frac{\dot{M}}{M_{\odot}
\text{yr}^{-1}} \right]^{1/4} \left[ \frac{M}{10^9 M_{\odot}} \right]^{-1/2},
	\end{equation}
	\begin{equation} \label{eq:nulnupeak}
		\frac{\nu_{\rm p} L_{\nu_{\rm p}}}{\text{[erg/s]}} = \mathcal{B} 
\left[ \frac{\dot{M}}{M_{\odot} \text{yr}^{-1}} \right] \cos \theta,
	\end{equation}

\noindent where Log$\mathcal{A} = 15.25$, Log$\mathcal{B}=45.66$. With a fixed inclination angle, these simple analytic expressions can be used to infer univocally the black hole mass $M$ and the accretion rate $\dot{M}$. Once we know $\nu_{\rm p}$ and $\nu_{\rm p} L_{\nu_{\rm p}}$, we aim to build an analogous analytic scaling for the KERRBB case.

\begin{figure}
\centering
\hskip -0.2 cm
\includegraphics[width=0.495\textwidth]{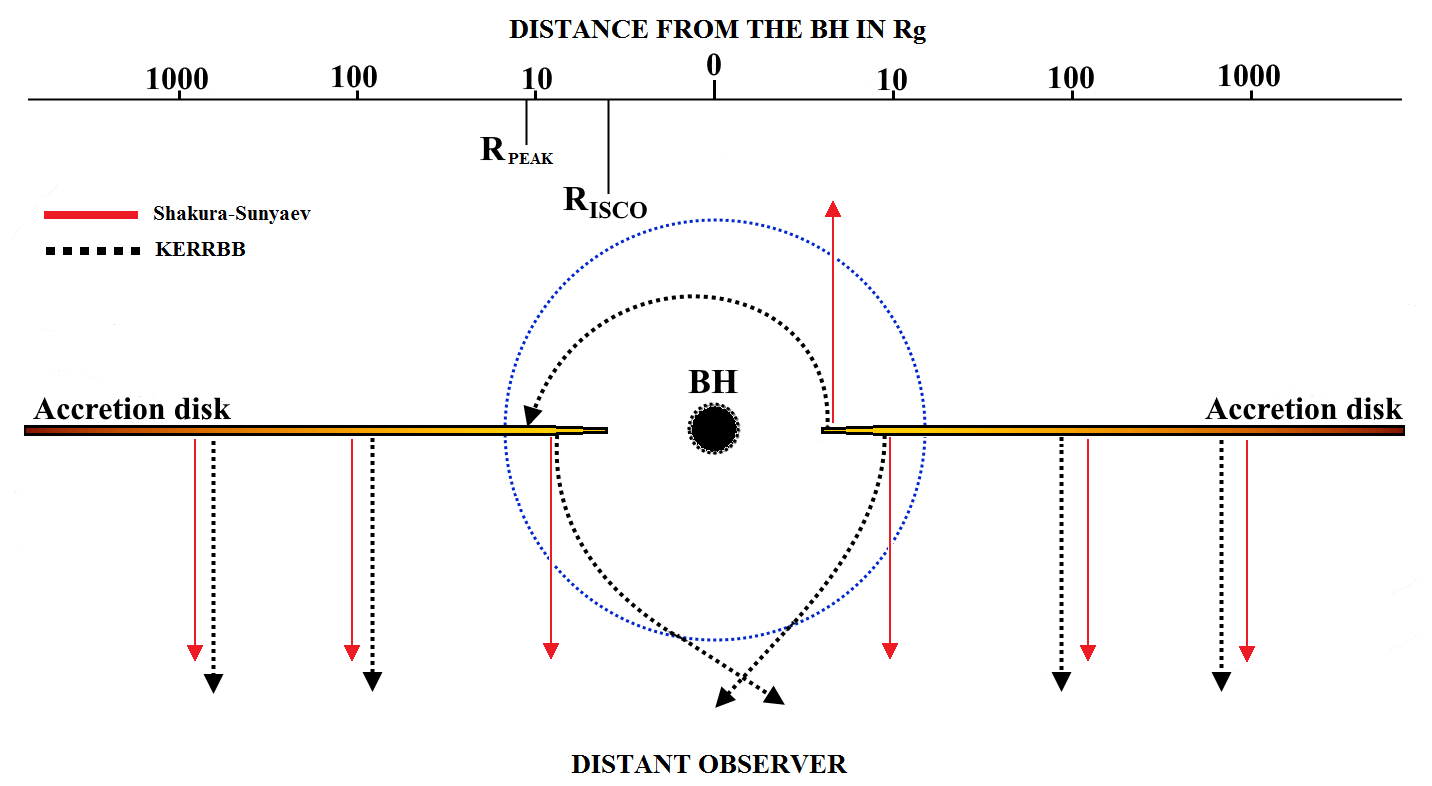}
\caption{Schematic view of the photon emission pattern of an accretion disk around a black hole. A distant observer is at the bottom of the figure and sees the disk face-on. For the classical SS model, photon trajectories (red lines) go straight to a distant observer; they are not affected by any relativistic effect. For the KERRBB model, photon trajectories (dotted black lines) are affected by light-bending and change their paths towards different directions with respect to the SS case (the region where light-bending is stronger is inside the dashed blue circle). Along with a lower efficiency, this explains why the KERRBB with the same spin ($a = 0$) is dimmer than the SS model when observed face-on. The photons emitted in the outer region of the disk are not affected by relativistic effects since they are far from the black hole influence, hence the SS model and the KERRBB model are similar. If the black hole is maximally spinning ($a = 0.9982$), the disk moves closer to the black hole and relativistic effects become stronger, which along with the larger efficiency make the disk brighter than the previous cases, even if the disk is edge-on with respect the observer. The top axis is the logarithmic distance from the central black hole in units of Rg: $R_{\rm ISCO}$ and radius $R_{\rm PEAK}$ (radius at which the SS emitted flux is maximized) are shown as well.} 
\label{fig:figurabella}
\end{figure}

The KERRBB model is a numerical code that describes the emission from a thin, steady state, general relativistic accretion disk around a rotating black hole. The local emission is assumed to be a blackbody. It was developed by \citet{Lietal} and is an extension of a previous relativistic model called GRAD (\citealt{Hanawa}, \citealt{Ebietal}), which assumes a nonrotating black hole. KERRBB is the most complete public code for fitting accretion disk spectra because it takes into account all the relativistic effects (i.e. Doppler beaming, gravitational redshift, light-bending, self-irradiation, Li05) and also the effects related to the black hole spin $a$ that determines the inner radius of the disk, the innermost stable circular orbit (ISCO). This radius ($R_{\rm ISCO}$) controls the radiative efficiency of the system (see also \S 3 and Appendix D in Li05). The flux coming from each annulus of the disk, emits like a diluted blackbody with the hardening factor $f_{\rm col} = T_{\rm col} / T_{\rm eff}$, where $T_{\rm eff} (R)$ is the effective temperature of the disk given by the Stefan-Boltzmann law, and $T_{\rm col}$ is the color temperature. The flux also depends on the radius $R$ and the dimensionless black hole spin $a$ (the relativistic flux is fully described in \citealt{PageTho}). The total disk luminosity is a function of the adimensional spin $a$ because the efficiency is related to the BH angular momentum: $\eta = 0.038$ for $a = -1$, $\eta = 0.057$ for $a = 0$, $\eta = 0.324$ for $a = +0.9982$; for $a = +1$, we formally have $\eta = 0.4226$, but this spin value can be reached only by ignoring the capture of radiation by the black hole (\citealt{Thorn74}).

In this work, we ignored the Comptonization process described by the hardening factor $f_{\rm col}$ since KERRBB modifies the whole spectrum. The assumption of a color temperature $T_{\rm col}$ higher than the effective local temperature $T_{\rm eff}$ could be true if, for instance, a hot corona is located above and below an otherwise standard accretion disk. In this case, the Comptonization process occurring in the corona can be mimicked by assuming $T_{\rm col} > T_{\rm eff}$. Although in an AGN the hot corona is present and active, it is so only in the inner regions of the accretion disk, while it is unimportant at larger radii. Hence the KERRBB model must account for this effect only at small radii and the corresponding X--ray emission cannot be approximated by a higher temperature of the disk but must be treated as a separate process. Therefore, in this work, we set the hardening factor $f_{\rm col} = 1$ and do not account for the Comptonization process. 

\begin{figure*} 
\centering
\includegraphics[width=0.47\textwidth]{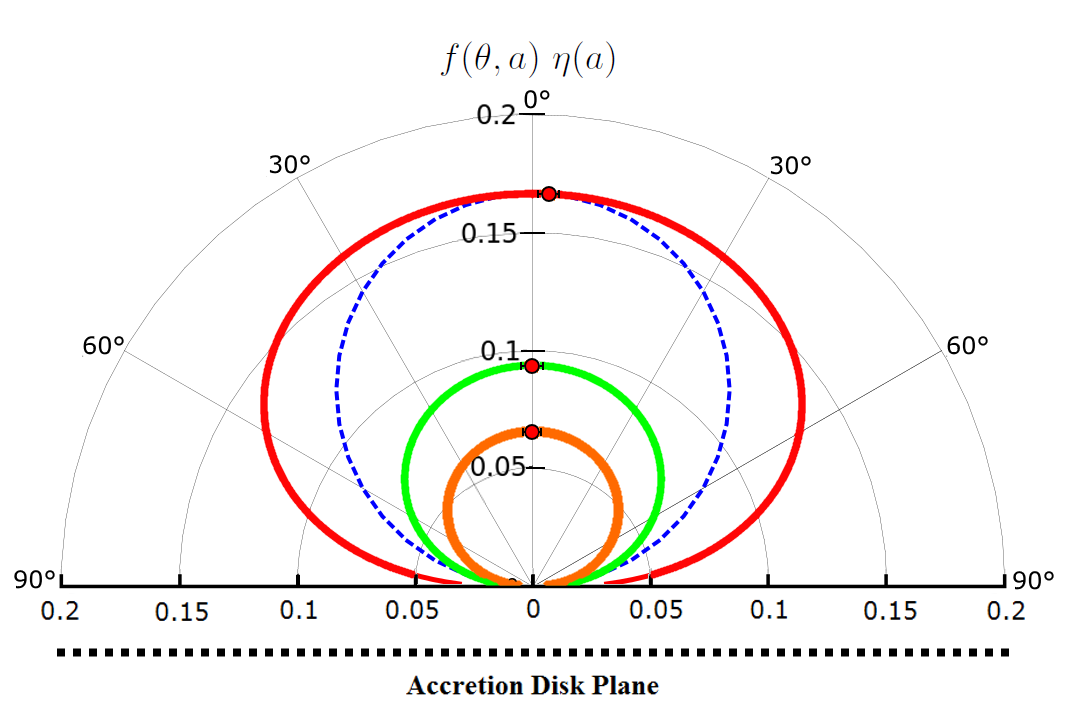} \qquad \includegraphics[width=0.48\textwidth]{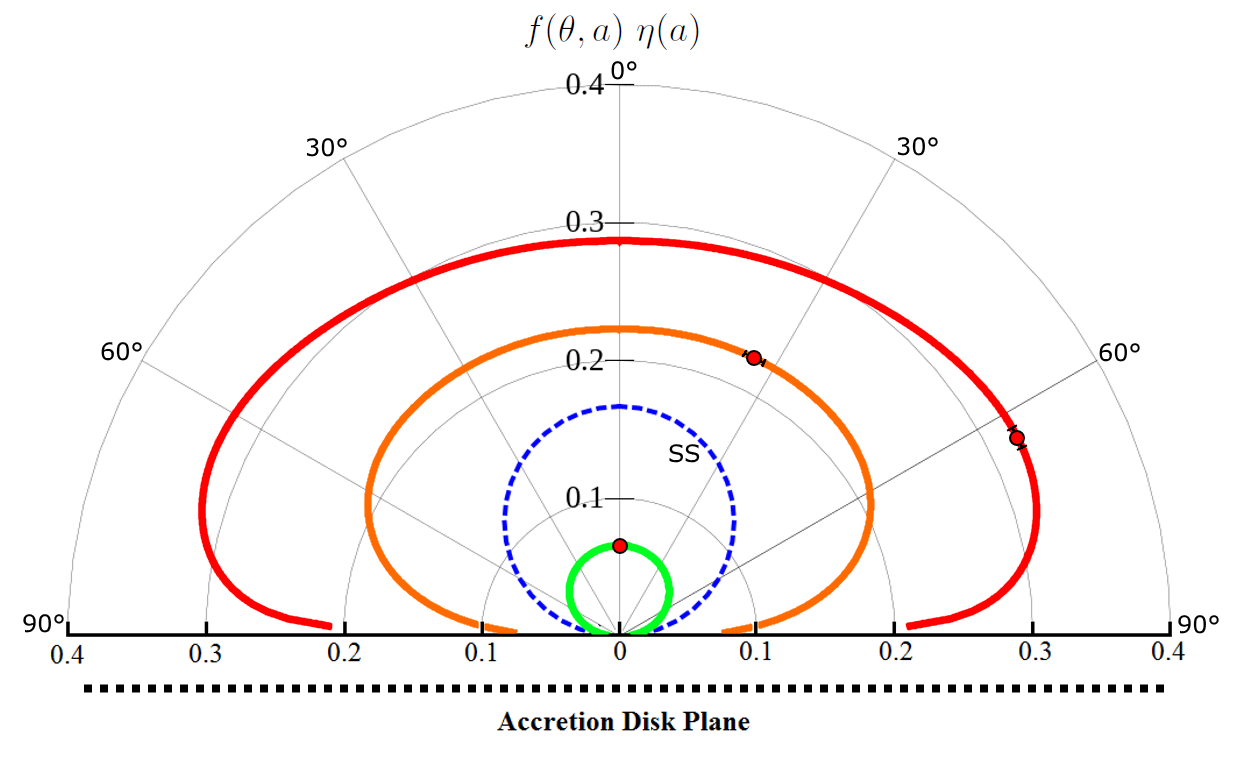}
\caption{Emission pattern for different models. The radial axis is the observed efficiency, given by $\eta_{\rm obs} \equiv f(\theta,a) \eta(a)$, i.e., the observed luminosity $L^{\rm obs}_{\rm d}$ normalized on $\dot{M} c^2$ (the accretion rate is the same for each model). Left panel: Classical nonrelativistic SS model (dashed blue line) compared with $a=-1$ (orange line), $a = 0$ (green line) and $a = 0.797$ (red line) KERRBB patterns. For the SS model ($\eta_{\rm SS} \sim 0.083$), the observed efficiency is $\eta_{\rm obs} = 2 \cos \theta \eta_{\rm SS} \approx 0.17 \cos \theta$. The $a = 0.797$ case describes the most similar KERRBB model to a SS with the same parameters (mass and accretion rate) at $\theta = 0^{\circ}$. At different viewing angles, the KERRBB model is brighter. For the SS model, the emission has a circular pattern while for KERRBB, the pattern is ``warped'' by the relativistic effects. Right panel: SS model (dashed blue line) compared with $a = -1$ (green line), $a = 0.95$ (orange line) and $a = 0.9982$ (red line) KERRBB patterns. For the extreme KERRBB, the emission is strongly modified with respect to the others and the observed disk luminosity is larger for larger angles. Red dots indicate the angle at which the KERRBB observed disk luminosity is maximized, obtained through Eq. \eqref{eq:deriv}: for $a \lsim 0.8$, $\theta_{\rm max} \sim 2^{\circ}$; for $a = 0.95$, $\theta_{\rm max} \sim 25^{\circ}$; for $a = 0.9982$, $\theta_{\rm max} \sim 64^{\circ}$ (see Table \ref{tab:angmax} for other spin values).}     
\label{fig:grid1}
\end{figure*}


\section{Analytic approximation of KERRBB disk emission} \label{sec:empat}

\subsection{Disk bolometric luminosity} \label{subsec:bollum}

\begin{figure}
\centering
\includegraphics[width=0.48\textwidth]{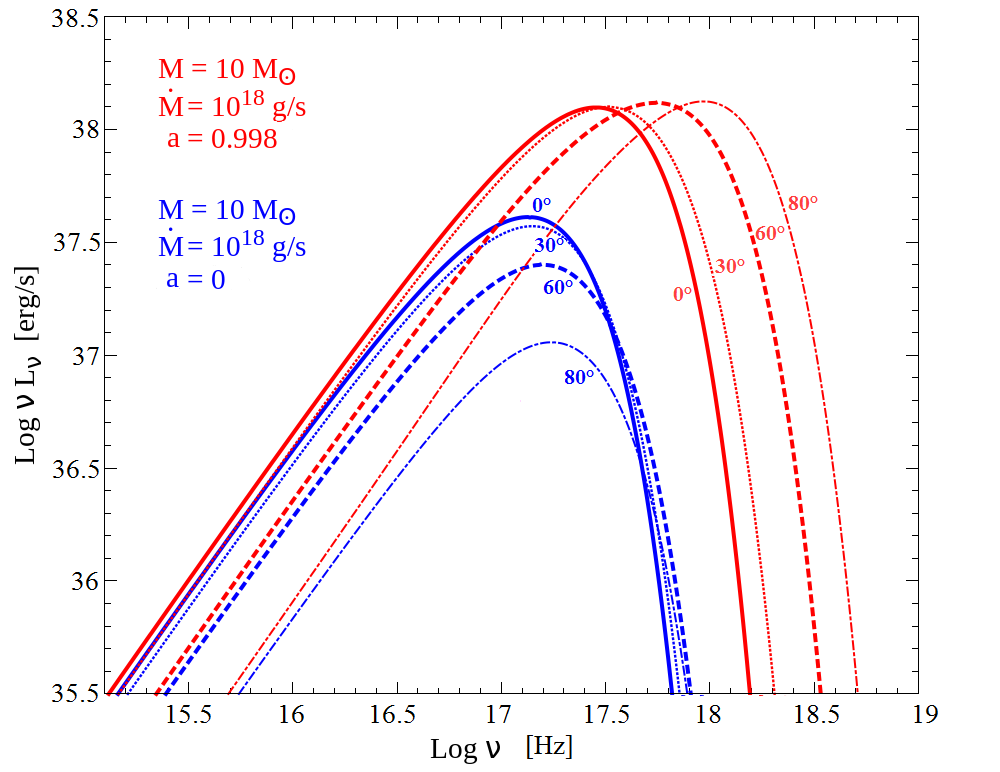}
\caption{
KERRBB model behavior at different angles and spins. All models have the same $M$ and $\dot{M}$. When $a = 0$ (blue lines), the disk luminosity decreases as the viewing angle increases in a way similar to $\cos \theta$. When the spin is maximal ($a = 0.9982$, red lines), the light-bending effect, along with Doppler beaming and gravitational redshift are so strong that the luminosity is almost the same, even at large angles. In this case, the trajectories of the energetic photons, produced close to the black hole horizon, are bent in all directions making the radiation intensity almost the same at all viewing angles. For a fixed viewing angle, the low-frequency part of the spectrum is almost the same for both $a=0$ and $a=0.9982$; this emission is related to the outer region of the disk where relativistic effects are not as important.   
\label{fig:spinang}}
\end{figure}

As shown in \S \ref{sec:basic}, for the classical nonrelativistic SS model, Eq. \eqref{SSobslum} describes the bolometric luminosity observed at an angle $\theta$, related to the total disk luminosity $L_{\rm d} = \eta \dot{M} c^2$. Equation \eqref{SSobslum} is no longer valid in the general relativistic case (\citealt{Cunnin}). In fact, relativistic effects modify not only the size of the inner radius of the disk, but also the pattern of the emitted photons: Fig. \ref{fig:figurabella} compares schematically the photon paths in the cases of SS and $a = 0$ KERRBB. In this latter case, light-bending plays a crucial role because it bends the photon trajectories towards different directions with respect to the normal of the disk. This effect, along with a lower efficiency $\eta$, explains why the $a = 0$ KERRBB model is dimmer than the SS one, when observed face-on. If the black hole is maximally spinning, the disk moves closer to the black hole and relativistic effects become stronger. Therefore, the efficiency is larger and the relativistic effects stronger, so that the disk is bright even for edge-on observers (see also Fig. \ref{fig:spinang}). Thus, the emission pattern depends rather strongly on the black hole spin $a$. Our task is to find an analytic expression for the pattern. To this end, we wrote the observed KERRBB bolometric disk luminosity as
	\begin{equation} \label{eq:kerrlum}
		L^{\rm obs}_{\rm Kerr} (\theta, a) = f(\theta, a) \eta(a) \dot{M} c^2.
	\end{equation}

\noindent This equation must satisfy
	\begin{equation} \label{eq:kerrlum2}
		\frac{1}{4 \pi} \int_{\Omega} L^{\rm obs}_{\rm Kerr} (\theta, a) d \Omega = \eta(a) \dot{M} c^2.
	\end{equation}

\noindent Using Eq. \ref{eq:kerrlum}, this implies
	\begin{equation} \label{eq:normalization}
		{\frac{1}{4 \pi}}\,2\times  \int_{0}^{\pi/2} f(\theta, a) d \Omega = 1  \to
\int_{0}^{\pi/2} f(\theta, a) \sin \theta d \theta = 1.
	\end{equation}

First, we consider the bolometric luminosity of the spectra given by the KERRBB code as a function of the viewing angle $\theta$, for given values of $\dot{M}$, $M$, $a$. In this way, we derive the emission pattern numerically. Then we search for an analytic expression that could interpolate the numerical results with good accuracy (see Fig. \ref{fig:intang} in the Appendix). We assume that the first term of the analytic expression is $\cos \theta$, for an easy comparison with Eq. \eqref{SSobslum}. Therefore, considering Eq. \eqref{eq:kerrlum}, we assume
	\begin{equation} \label{eq:kerrlum3}
		L^{\rm obs}_{\rm Kerr} (\theta, a) = 
\underbrace{[ \cos \theta \cdot (\text{terms functions of $\theta$})]}_{= f(\theta, a)} \ \eta(a) \dot{M} c^2.
	\end{equation}

\noindent A good match is found using the following functional form:
	\begin{equation} \label{eq:kerrfinale}
		L^{\rm obs}_{\rm Kerr} (\theta, a) = \underbrace{A \cos \theta [1 - (\sin \theta)^C]^B 
[1 - E (\sin \theta)^F]^D}_{= f(\theta, a)} \ \eta(a) \dot{M} c^2.
	\end{equation}

\noindent In this form, $f(\theta, a)$ differs from the numerical result by less than $\lsim 1 \%$ (see Sect. \ref{POLY} in the Appendix for the polynomial function approximation). All the parameters $A$, $B$, $C$, $D$, $E$ and $F$ are functions of the spin $a$. We then repeat this analysis using different spin values. With this approach, we study how the values of these parameters change by changing the spin. Again, we search for an analytical function that interpolates the numerical results (see Fig. \ref{fig:intang2} in the Appendix). We find that a good representation is given by the following function:
	\begin{eqnarray} \label{eq:kerrpara}
		\mathcal{F}(a) &=& \alpha + \beta x_1 + \gamma x_1^2 + \delta x_1^3 + \epsilon x_1^4 + \iota x_1^5 + \kappa x_1^6 
		\nonumber \\
		x_1 &=& \log(1-a)
	\end{eqnarray}

The values of $\alpha$, $\beta$, $\gamma$, $\delta$, $\epsilon$, $\iota$, $\kappa$ are listed in Table \ref{tab:parsping}. The accuracy of Eq. \ref{eq:kerrpara} is $\sim 1 \%$. For any spin value, the values in Table \ref{tab:parsping} can be used to calculate the parameters $A$,$B$,... $F$ of Eq. \ref{eq:kerrfinale}. The observed luminosity $L^{\rm obs}_{\rm Kerr} (\theta, a)$ can therefore be obtained for any disk inclination angle in the range $[0^{\circ}:85^{\circ}]$ (i.e., the angle range of KERRBB). As an example, in the case with $a = 0$ ($\eta = 0.057$), the function that interpolates the observed disk luminosity for different angles $\theta$ is
	\begin{equation} \label{eq:kerrex}
		\frac{L^{\rm obs}_{\rm Kerr} (\theta, a=0)}{\dot{M} c^2} = 0.094 \frac{\cos \theta [1 - 0.958 (\sin \theta)^{6.675}]^{0.121}}{[1 - (\sin \theta)^{1.918}]^{0.202} }
	\end{equation}

\subsection{Pattern} \label{subsec:pat}

Through Eq. \eqref{eq:kerrlum}, whose explicit form is Eq. \eqref{eq:kerrfinale}, it is possible to study the emission pattern of the Kerr black hole accretion disk. It is interesting to compare it with the pattern of the SS model (i.e. with its $\cos\theta$ pattern in Eq. \eqref{SSobslum}). Relativistic effects radically change the $\theta$-dependence. The observed luminosity is not maximized at $0^{\circ}$ any longer, but at a different angle $\theta_{\rm max}$. Setting the derivative of Eq. \eqref{eq:kerrfinale} with respect to $\theta$ to zero, we find
	\begin{equation} \label{eq:deriv}
		\begin{split}
		&\ [1 - (\sin \theta)^C] \cdot [1 - E (\sin \theta)^F] \\
		&+\ C B  (\sin \theta)^{C-2} \cos^2 \theta\ [1 - E (\sin \theta)^F] \\
		&+\ FED (\sin \theta)^{F-2} \cos^2 \theta\ [1 - (\sin \theta)^C]  = 0
		\end{split}
	\end{equation}

\noindent with a mean uncertainty of $\sim 5^{\circ}$. The angle $\theta_{\rm max}$ increases for increasing spin. For spins $a \lsim 0.9$, $\theta_{\rm max}$ is less than $\sim 10^{\circ}$, i.e., relativistic effects do not strongly modify the emission pattern. For $a > 0.9$, relativistic effects become stronger: for example, $\theta_{\rm max} \simeq 25^{\circ}$ for $a = 0.95$, and $\simeq 65^{\circ}$ for $a = 0.9982$ (see Table \ref{tab:angmax} for other spin values). We note that Eq. \eqref{eq:deriv} is equal to zero also for $\theta = 90^{\circ}$; this solution must not be considered because Eq. \eqref{eq:kerrfinale} is defined in the interval $[0^{\circ}:85^{\circ}]$.

\begin{figure*}
\centering
\hskip -0.2 cm
\includegraphics[width=0.47\textwidth]{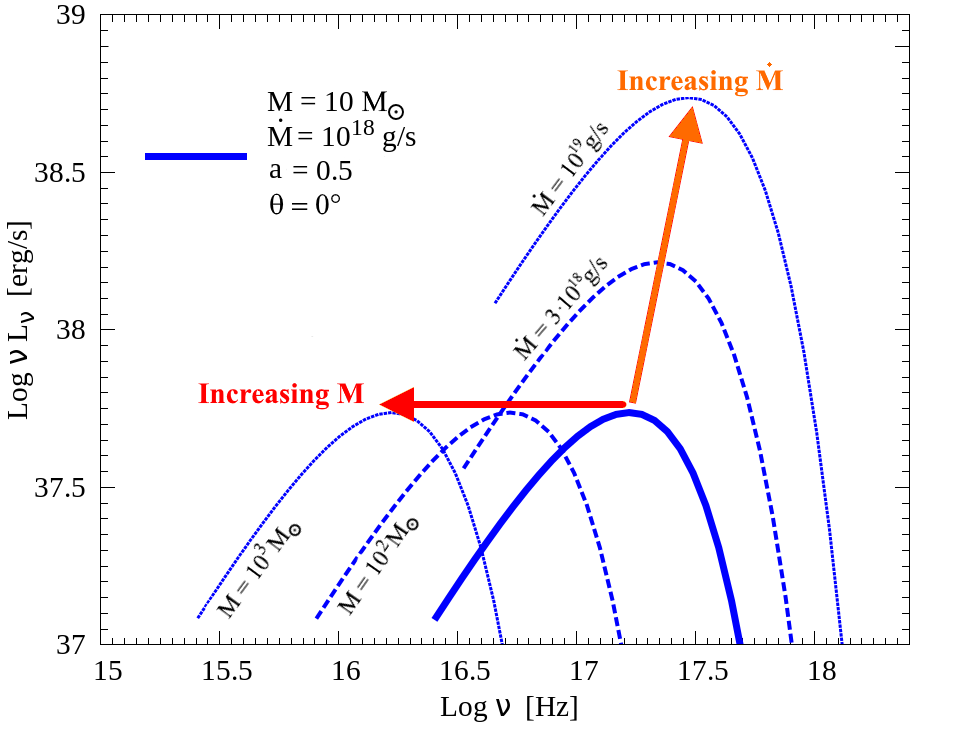} \qquad \includegraphics[width=0.45\textwidth]{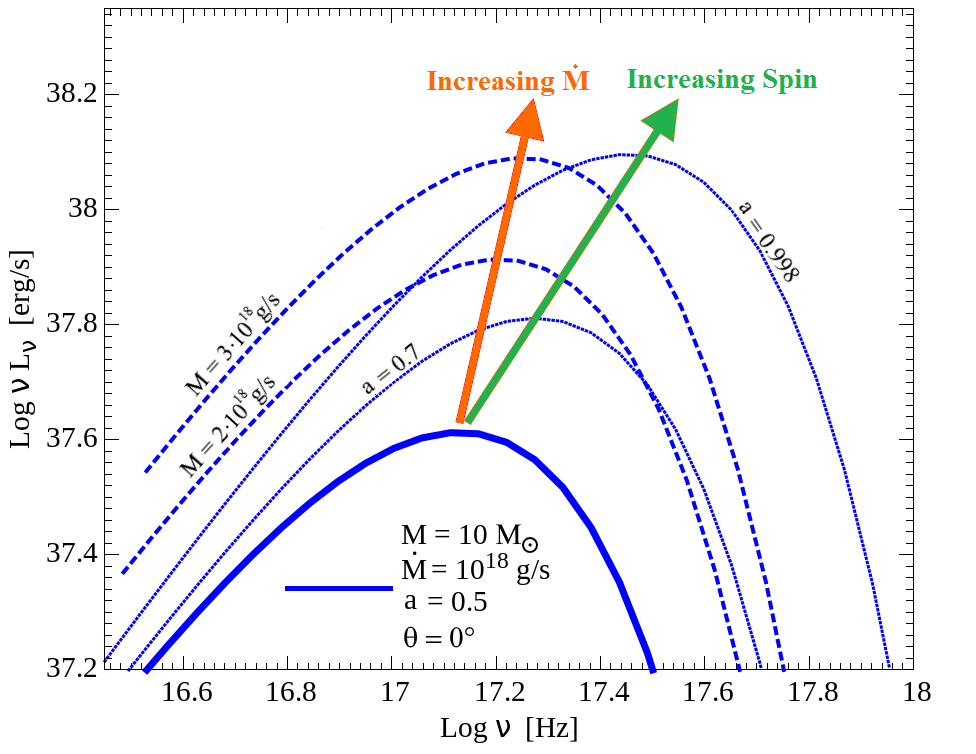}
\caption{
KERRBB accretion disk spectra with different masses, accretion rates and black hole spins. Left panel: The initial $10 M_{\odot}$ spectrum (blue thick solid line) moves horizontally (i.e., in frequency) when the black hole mass changes (red arrow), and diagonally (i.e., in both frequency and luminosity) when the accretion rate changes (orange arrow). The position of the peak in frequency $\nu$ and luminosity $\nu L_{\nu}$ for a given spin value follows the same shifting equations found by \citet{Caldero}. Right panel: When $\dot M$ increases, the $10 M_{\odot}$ spectrum peak also increases (orange arrow), but less than the increase related to the increase of the spin (green arrow). When $a$ changes the low-frequency part of the spectrum remains constant, but this occurs at frequencies not visible in the figure. The low-frequency flux is produced by the outer regions of the disk, unaffected by the black hole spin. About this effect, see also Fig. \eqref{fig:spinang}.
} 
\label{fig4}
\end{figure*}
 
Figure \ref{fig:grid1} shows the emission pattern for different models; the radial axis is the observed efficiency $\eta_{\rm obs} \equiv f(\theta,a) \eta(a)$ for different viewing angles. In the SS case (dashed blue line), $\eta_{\rm obs} = 2 \cos \theta \eta_{\rm SS} \approx 0.17 \cos \theta$. The left panel compares the KERRBB patterns for $a = -1$ (orange line), $0$ (green line) and $0.797$ (red line). We note that the SS model, when observed at $\theta = 0^{\circ}$ is well approximated by the KERRBB model with $a = 0.797$. Indeed the efficiency of a nonspinning black hole ($\eta=0.057$) is lower than the Newtonian one, generally assumed for a SS model ($\eta_{\rm SS} \sim 0.083$). Therefore, to match the SS luminosity a larger efficiency is needed, and hence a larger spin. Furthermore, at $\theta = 0^{\circ}$ light-bending decreases the emission in the relativistic case. An even larger spin is thus required to match the SS emission. For the same reason, at larger angles the KERRBB model is brighter. The right panel compares the SS model with the $a = -1$ (green line), $0.95$ (orange line) and $0.9982$ (red line) KERRBB patterns. As expected, for larger spins the KERRBB emission is strongly modified and the observed disk luminosity is larger for larger angles. Red dots indicate the angles at which the observed disk luminosity is maximized. The strong modification of the spectrum emission due to the combination of viewing angle and spin is also visualized in Fig. \ref{fig:spinang}: the blue spectra correspond to $a = 0$, the red spectra to $a = 0.9982$ ($M$ and $\dot{M}$ are the same). In the $a = 0$ case, relativistic effects are very weak and the spectrum almost follows the $\cos\theta$--law. Instead, in the $a = 0.9982$ cases, $L^{\rm obs}_{\rm Kerr} (\theta)$ is almost constant, even for the largest viewing angles. This is due to the combination of different relativistic effects (Doppler beaming, gravitational redshift and light-bending) along with the black hole spin: the trajectories of the energetic photons coming from the innermost region of the disk (which are very close to the horizon for $a \to 1$) are bent in all directions and the intensity of radiation is almost the same at all viewing angles.    

\begin{table} 
\centering
\footnotesize
\begin{tabular}{lllllllllll}
\hline
Spin value [$a$] & $\theta_{max}$ & Spin value [$a$] & $\theta_{max}$\\
\hline
$0.9982$ & $65^{\circ}$ & $0.6$ & $0.4^{\circ}$ \\
$0.98$ & $46^{\circ}$ & $0.5$ & $0.2^{\circ}$ \\
$0.95$ & $25^{\circ}$ & $0.4$ & $0.1^{\circ}$ \\
$0.94$ & $19^{\circ}$ & $0.3$ & $0.03^{\circ}$ \\
$0.92$ & $13^{\circ}$ & $0.2$ & $0.01^{\circ}$ \\ 
$0.9$ & $9^{\circ}$ & $0.1$ & $\sim 0^{\circ}$ \\ 
$0.8$ & $2^{\circ}$ & $0$ & $\sim 0^{\circ}$ \\
$0.7$ & $1^{\circ}$ & $-1$ & $\sim 0^{\circ}$\\
\hline \\
\end{tabular}
\caption{
Values of the viewing angle $\theta_{\rm max}$ at which the disk bolometric luminosity is maximized, for different spins, obtained through Eq. \eqref{eq:deriv}. The mean scattering from the real value of $\theta_{\rm max}$ is $\sim 5^{\circ}$. For spins $a \lsim 0.9$, $\theta_{\rm max}$ is less than $\sim 10^{\circ}$, i.e., relativistic effects do not strongly modify the emission pattern. For $a > 0.9$, relativistic effects become stronger: we note the extreme effect at spin $a = 0.9982$ with respect to the other models. These results show the deviation from the $\cos \theta$--law at different spins.
} 
\label{tab:angmax}
\end{table}

\begin{table*} 
\centering
\footnotesize
\begin{tabular}{lllllllllll}
\hline
Parameters & $\alpha_g$ & $\beta_g$ & $\gamma_g$ & $\delta_g$ & $\epsilon_g$ & $\zeta_g$ & $\iota_g$ \\ 
\hline   
$g_1$ & 1001.3894 & -0.061735 & -381.64942 & 8282.077258 & -40453.436 & 66860.08872 & -34974.1536 \\
$g_2$ & 2003.6451 & -0.166612 & -737.3402 & 16310.0596 & -80127.1436 & 132803.23932 & -69584.31533 \\
\hline \\
\end{tabular}
\caption{Parameters of the functions $g_1$ and $g_2$ in Eq. \eqref{eq:nupeak2} and Eq. \eqref{eq:nulnupeak2}, written as a general functional \eqref{eq:functional2}, in the case with the viewing angle $0^{\circ}$. Using KERRBB results, it is possible to find similar equations for the cases with the disk inclination angle $> 0^{\circ}$.   
\label{ap:tab1}}
\end{table*}


\subsection{Scaling with black hole mass and accretion rate} \label{scale}

In the SS model the position of the peak frequency $\nu_{\rm p}$ and the peak luminosity $\nu_{\rm p} L_{\nu_{\rm p}}$ scale with the mass and the accretion rate according to Eq. \eqref{eq:nupeak} and Eq. \eqref{eq:nulnupeak}. It is possible to find analogous scalings for the KERRBB case. Fixing $a$ and $\theta$ in the KERRBB model, the peak position scales following Eq. \eqref{eq:nupeak} and Eq. \eqref{eq:nulnupeak}, as in SS (Fig. \ref{fig4}). In fact, the assumption of a local blackbody emission leads to a similar temperature profile (\citealt{NovTho}, \citealt{PageTho}). Therefore, we write the general KERRBB scalings as
	\begin{equation} \label{eq:nupeak2}
		\frac{\nu_{\rm p}}{\text{[Hz]}} = \mathcal{A} \left[ \frac{\dot{M}}{M_{\odot} \text{yr}^{-1}} \right]^{1/4} 
\left[\frac{M}{10^9 M_{\odot}} \right]^{-1/2} g_1(a, \theta),
	\end{equation}
	\begin{equation} \label{eq:nulnupeak2}
		\frac{\nu_{\rm p} L_{\nu_{\rm p}}}{\text{[erg/s]}} = \mathcal{B} \left[ \frac{\dot{M}}{M_{\odot} \text{yr}^{-1}} \right] g_2(a, \theta) \cos \theta,
	\end{equation}

\noindent where the functions $g_1$ and $g_2$ describe the dependencies on the black hole spin $a$ and the viewing angle $\theta$. 

\begin{figure}
\centering
\hskip -0.2 cm
\includegraphics[width=0.49\textwidth]{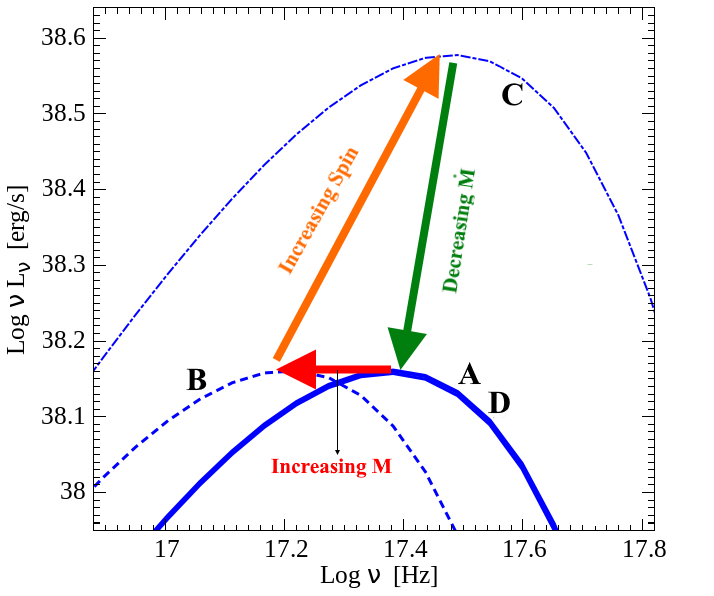}
\caption{
Emission peak of the KERRBB spectrum in the same position for different values of the black hole mass $M$, accretion rate $\dot M$ and spin $a$ which shows that there is a family of solutions. Consider spectrum A (thick solid blue line): if $M$ increases at constant $\dot M$ and spin, spectrum B is obtained. Now, if the spin increases, but $M$ and $\dot M$ are constant, spectrum C is obtained because a larger spin corresponds to a larger efficiency. Finally, if $\dot M$ decreases, spectrum D is obtained, which is almost exactly equal to the initial spectrum A (they differ slightly in the high-frequency, exponential part). 
}
\label{fig:figura5b}
\end{figure}

Since in the following we apply our method to blazars, believed to have $\theta \sim 0^{\circ}$, here we are focusing on this case. Therefore, we derived $g_1(a, \theta=0^{\circ})$ and $g_2(a, \theta=0^{\circ})$:
	\begin{eqnarray} \label{eq:functional2}
		g_i(a, 0^{\circ}) &=& \alpha_g + \beta_g y_1 + \gamma_g y_2 + \delta_g y_3 + \epsilon_g y_4 +\zeta_g y_5 + \iota_g y_6
		\nonumber \\
		y_{\rm n} &\equiv& \log(n-a) \qquad \qquad \qquad i = 1, 2
	\end{eqnarray}

Table \ref{ap:tab1} gives the parameter values of the two functions. Equation \ref{eq:functional2} approximates numerical data better than $\sim 1 \%$. This process can be repeated also for any value of $\theta$, using the corresponding KERRBB results. 

It is important to note that the peak frequency and luminosity, described by Eq. \eqref{eq:nupeak2} and Eq. \eqref{eq:nulnupeak2}, are degenerate in mass, accretion rate and spin (if $\theta$ is fixed). In other words, the same spectrum can be fitted by a family of solutions, by changing $M$, $\dot{M}$ and $a$ in a proper way. Figure \ref{fig:figura5b} shows this degeneracy: starting from model A, by changing $M$, $\dot{M}$ and $a$ we can obtain spectrum D, which is equal to the initial spectrum A \footnote{Strictly speaking, models A and D are not exactly equal; there are very small differences in the high-frequency, exponential part of the spectra. For larger spin values, the exponential tail is brighter.}. The overlapping of two models can be done because, as shown in Fig. \ref{fig4} the accretion rate and the spin move the spectrum peak in different directions. 

\begin{figure}
\centering
\includegraphics[width=0.49\textwidth]{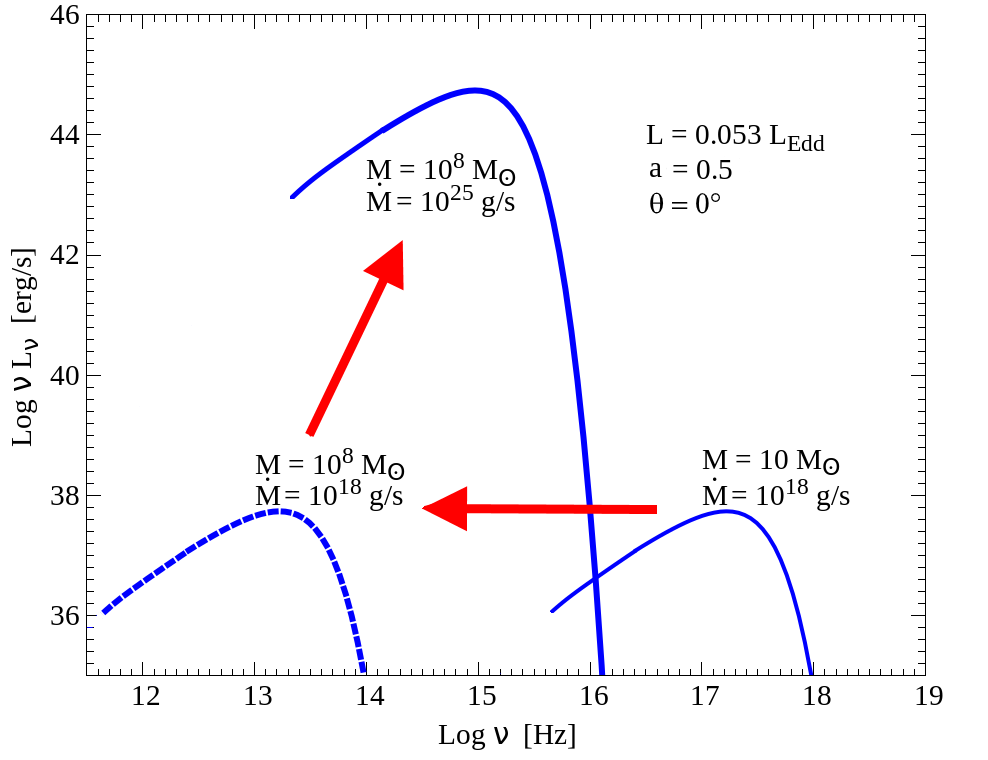}
\caption{
Scaling mass and accretion rate from a stellar black hole to a supermassive black hole. The initial KERRBB model (thin solid blue line) is first shifted in frequency by changing the mass value then in luminosity (and frequency) by changing the accretion disk value according to equations \eqref{eq:shift1}. In this way, it is possible to describe the accretion disk emission data related to supermassive black holes with KERRBB. The initial and the final spectra have the same Eddington ratio $L / L_{\rm Edd} = 0.053$. In the Appendix, equations \eqref{eq:shift1} are expressed as a function only of the black hole mass by considering that the Eddington ratio of the initial and the final models must be equal. The low-frequency part of the spectra is not shown in order to have a clearer figure.}
\label{fig:figura5a}
\end{figure}

Since the KERRBB spectrum moves on the Log$\nu$, Log$\nu L_{\nu}$ plane according to Eq. \eqref{eq:nupeak2} and Eq. \eqref{eq:nulnupeak2}, we can scale the spectrum from stellar to supermassive black hole masses with fixed spin and $\theta$. This procedure can be done if the emission processes from an accretion disk around a stellar black hole and a SMBH are the same. In our case, in both systems the emission is assumed to be a multicolor blackbody. Figure \ref{fig:figura5a} shows an example of our procedure: starting from a stellar black hole mass (thin solid blue line), the spectrum can be shifted in frequency and luminosity by quantities related to the initial and the final mass:
	\begin{equation} \label{eq:shift1}
		\frac{\nu_{\rm fin}}{\nu_{\rm in}} = \left[ \frac{\dot{M}_{\rm fin}}{\dot{M}_{\rm in}} \right]^{1/4} \sqrt{ \frac{M_{\rm in}}{M_{\rm fin}} } \qquad \qquad \qquad 
		\frac{\nu L_{\nu_{\rm fin}}}{\nu L_{\nu_{\rm in}}} = \frac{\dot{M}_{\rm fin}}{\dot{M}_{\rm in}}
	\end{equation}

In this way, it is possible to describe the accretion disk emission data related to supermassive black holes with KERRBB. In the Appendix, Eq. \eqref{eq:shift1} are expressed as a function only of the black hole mass. This is possible if the Eddington ratio is the same [Eq.~\eqref{eq:shift1apnew}].


\section{Family of KERRBB solutions} \label{KERRfamily}

Knowing the two observables $\nu_{\rm p}$ and $\nu_{\rm p}L_{\nu_{\rm p}}$, we can identify a family of solutions with $M$, $\dot{M}$, $a$ and $\theta$ linked by Eq. \eqref{eq:nupeak2} and Eq. \eqref{eq:nulnupeak2}. In specific cases, it is possible to have independent information on some of these parameters. For instance, for blazars we know that $\theta = 0^\circ$. In this case, if we know the spin we can estimate both $M$ and $\dot M$. If instead we have information on the mass, we can estimate $a$ and $\dot{M}$.

Figure \ref{Methspin} illustrates how $a$, $M$ and $\dot{M}$ are related in the case of $\theta=0^\circ$ for a disk emission peaking at Log $\nu_{\rm p} \approx 15.05$ and Log $\nu_{\rm p} L_{\nu_{\rm p}} \approx 44.86$. This figure shows the allowed solutions as a curve in the $\dot{M}$--$M$ plane. Each point of this curve is associated with a single spin value, as shown by the color-coding. This is the basis to observationally constrain the physical parameters of a blazar disk, as detailed in \S\ref{appl}. 

\begin{table} 
\centering
\footnotesize
\begin{tabular}{lllllllllll}
\hline
Spin $a$ & \hspace{3mm} $\eta$ & \hspace{3mm} Log $M_{\rm BH}$ & \hspace{2mm} Log $\dot{M}$ \\
\hline   
-1	   & \hspace{3mm} 0.038 & \hspace{3mm} 8.63	& \hspace{2mm} 0.39 \\
-0.8   & \hspace{3mm} 0.040 & \hspace{3mm} 8.66	& \hspace{2mm} 0.36 \\
-0.6   & \hspace{3mm} 0.043 & \hspace{3mm} 8.68	& \hspace{2mm} 0.34 \\
-0.4   & \hspace{3mm} 0.047 & \hspace{3mm} 8.71	& \hspace{2mm} 0.32 \\
-0.2   & \hspace{3mm} 0.051 & \hspace{3mm} 8.74	& \hspace{2mm} 0.30 \\
0	   & \hspace{3mm} 0.057 & \hspace{3mm} 8.78	& \hspace{2mm} 0.27 \\
0.2	   & \hspace{3mm} 0.064 & \hspace{3mm} 8.82	& \hspace{2mm} 0.24 \\
0.4	   & \hspace{3mm} 0.075 & \hspace{3mm} 8.87	& \hspace{2mm} 0.22 \\
0.6	   & \hspace{3mm} 0.091 & \hspace{3mm} 8.93	& \hspace{2mm} 0.19 \\
0.8	   & \hspace{3mm} 0.122 & \hspace{3mm} 9.02	& \hspace{2mm} 0.15 \\
0.9	   & \hspace{3mm} 0.156 & \hspace{3mm} 9.08	& \hspace{2mm} 0.13 \\
0.998  & \hspace{3mm} 0.321 & \hspace{3mm} 9.18	& \hspace{2mm} 0.09 \\
0.9999 & \hspace{3mm} 0.382 & \hspace{3mm} 9.20	& \hspace{2mm} 0.08 \\
\hline \\
\end{tabular}
\caption{Data of  Fig. \ref{Methspin}. 
KERRBB solutions for a disk with Log$\nu_{\rm p} \approx 15.05$ and Log$\nu_{\rm p} L_{\nu_{\rm p}} \approx 44.86$. Masses are in units of solar masses and accretion rate in units of solar masses per year.
\label{datafig45}}
\end{table}


\subsection{Constraining the black hole spin} \label{CostrSpin}

\begin{figure}
\centering
\hskip -0.2 cm
\includegraphics[width=0.49\textwidth]{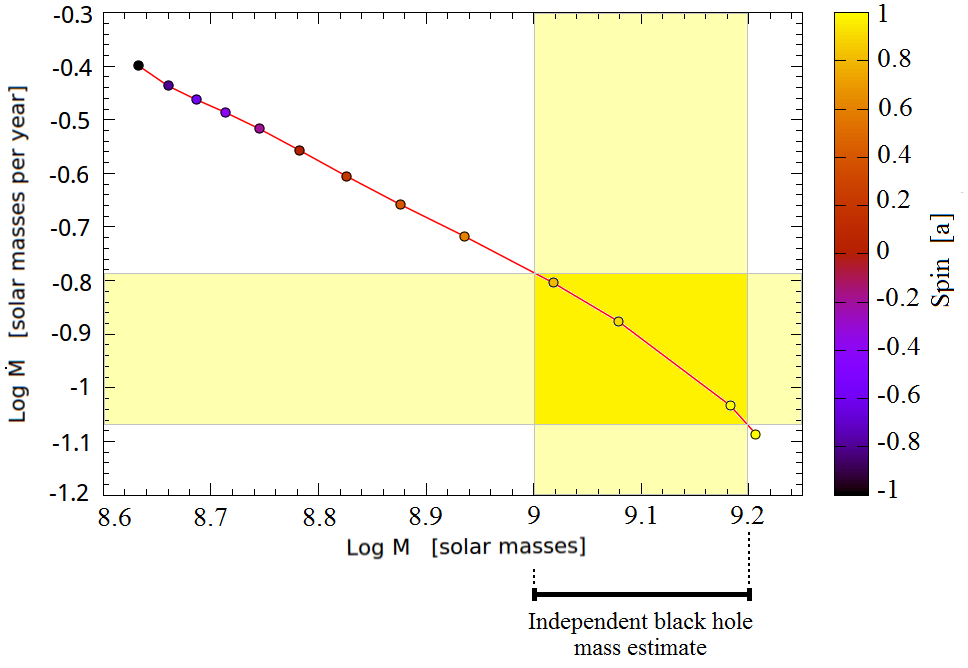} 
\caption{
KERRBB solutions for a spectrum with Log $\nu_{\rm p} \approx 15.05$ and Log $\nu_{\rm p} L_{\nu_{\rm p}} \approx 44.86$ (Table \ref{datafig45}). Black hole spins are constrained using an independent black hole mass estimate Log $M_{\rm BH, ind} = 9.1 \pm 0.1$. In this case, the range of the allowed black hole spins is $a = 0.93^{+0.07}_{-0.17}$.}
\label{Methspin}
\end{figure}

Let us assume that we are observing $\nu_{\rm p}$ and $\nu_{\rm p} L_{\nu_{\rm p}}$. If we have an independent estimate of the black hole mass with its uncertainty (e.g., through the virial method), we can constrain spin and accretion rate of the system. Figure \ref{Methspin} shows $\dot M$ and $a$ can be found. The vertical stripe corresponds to the independent mass estimate (with its uncertainty). This intercepts the curve of solutions and identifies a range of possible $\dot M$ and spins. Clearly, the narrower the range of black hole masses, the more precise the $\dot M$ and $a$ estimates. When no independent mass estimate is available, a rough limit can be given by assuming that the disk luminosity is sub-Eddington: this yields a lower limit on $M$ and $a$, and an upper limit on $\dot{M}$.

\subsection{Constraining the black hole mass} 

Parallel to the previous subsection, let us now suppose that we have an independent estimate on the black hole spin and its uncertainties. Figure \ref{Methspin} shows that this range corresponds to a section of the solution curve, allowing one to find a corresponding range of $M$ and $\dot M$. Differently from the black hole mass estimates, which are easily applicable and widely used, deriving the spin is more difficult. However, there is a general consensus that relativistic jets are generated by tapping the rotational energy of black holes with large spins. In the next section we will use this property to derive the black hole masses and accretion rates of two blazars.


\begin{figure*} 
\centering
\includegraphics[width=0.475\textwidth]{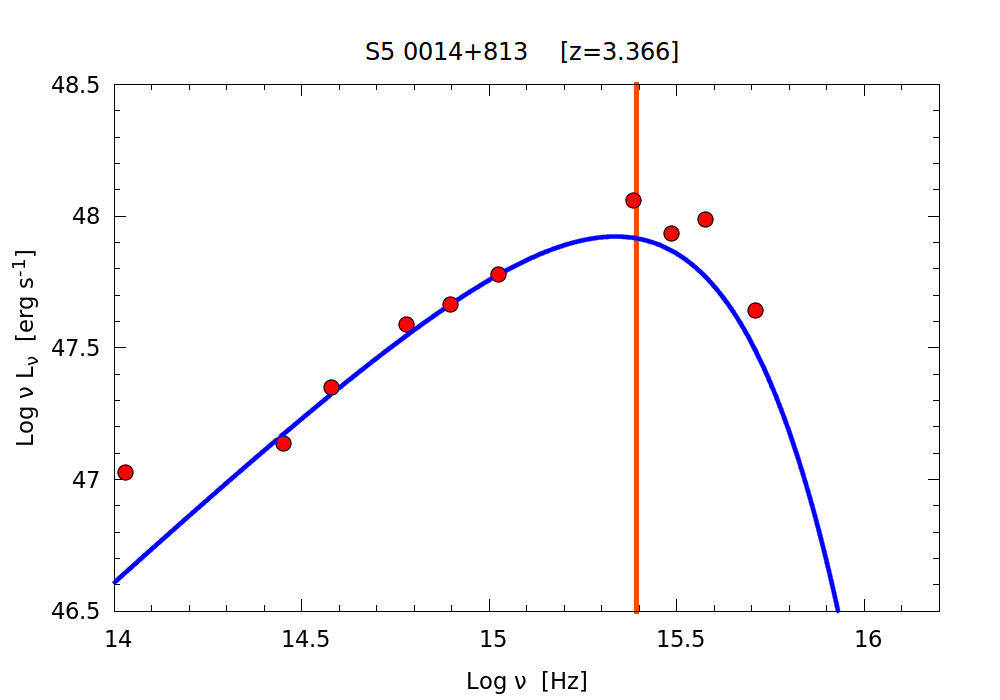} \qquad \includegraphics[width=0.475\textwidth]{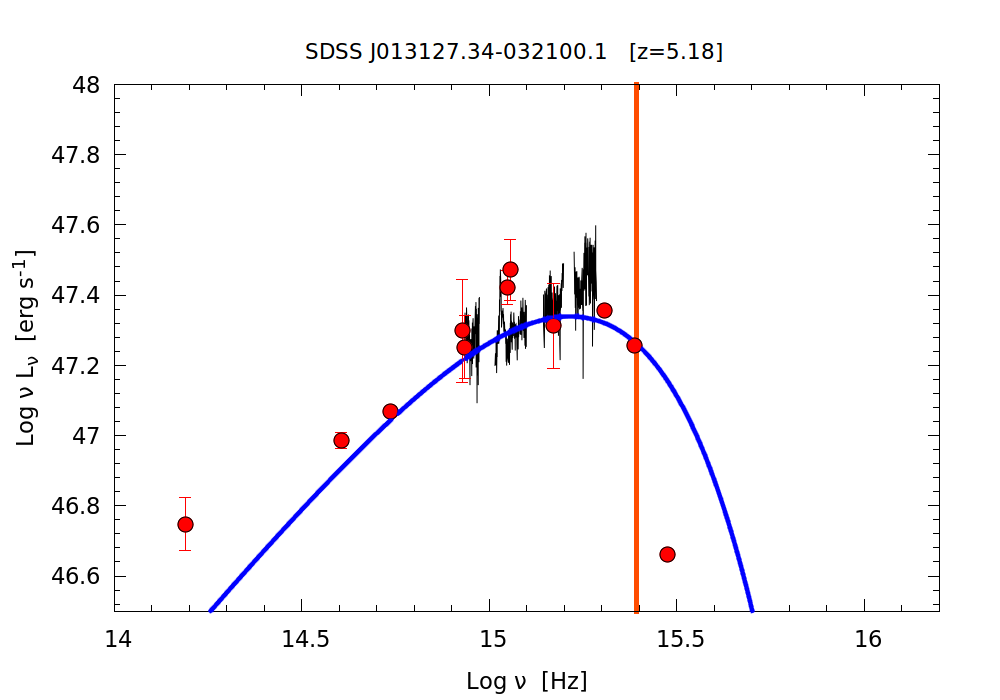}
\includegraphics[width=0.475\textwidth]{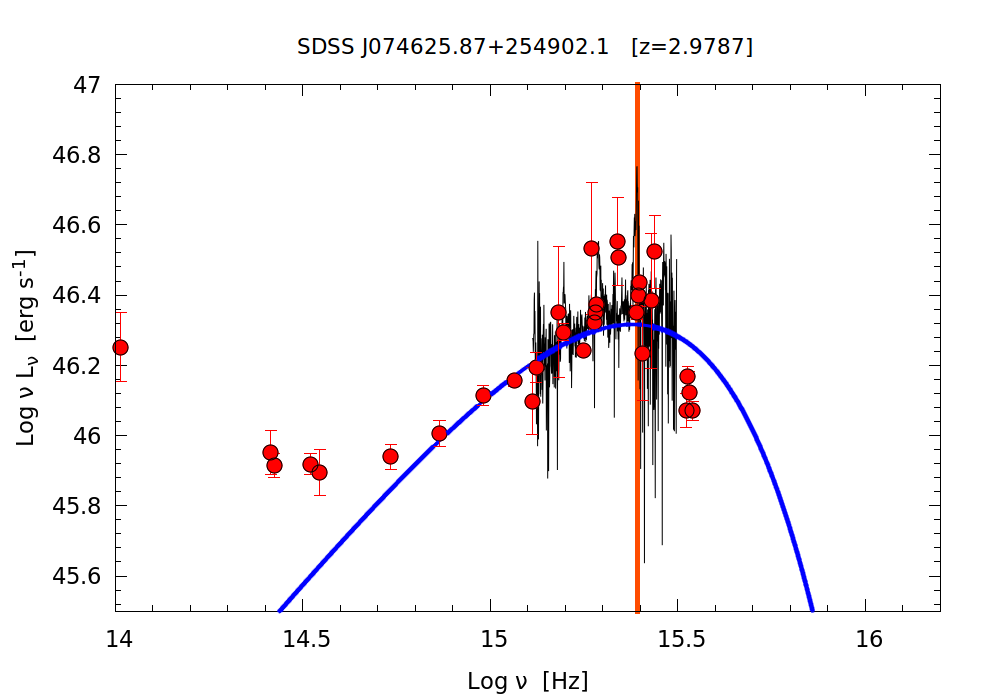} \qquad \includegraphics[width=0.475\textwidth]{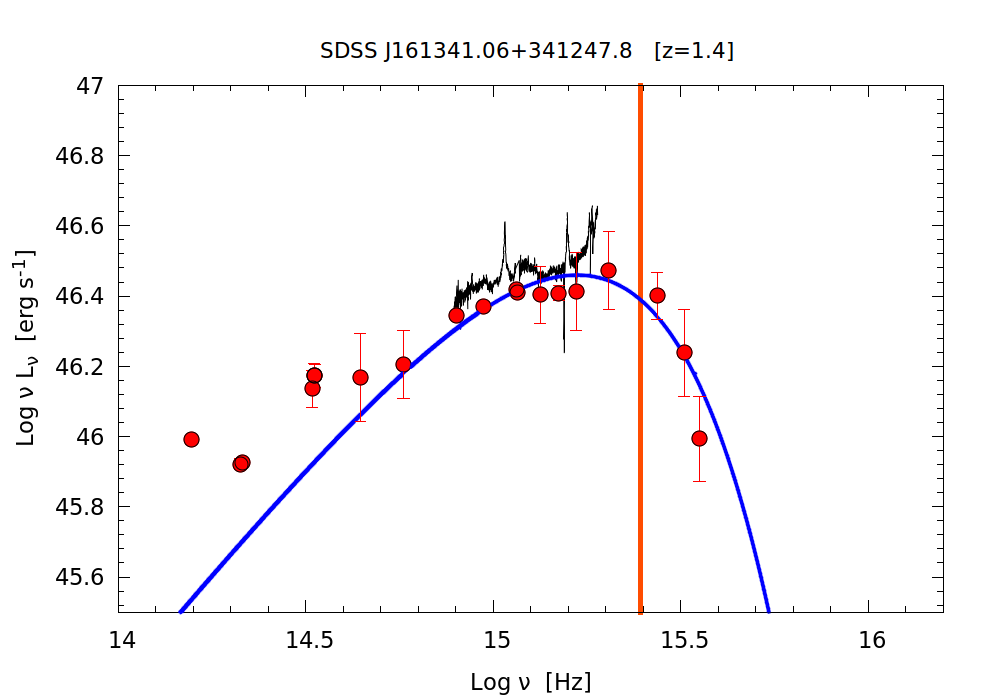}
\caption{Top left panel: SED of the blazar S5 0014+813 along with a KERRBB model (blue line). The model has $\nu_{\rm p} \approx 15.33$ and $\nu_{\rm p} L_{\nu_{\rm p}} \approx 47.92$. The SS mass is Log$M_{\rm SS} = 9.96$; the KERRBB mass limits, corresponding to $a = -1$ and $a = 0.9999$, are Log$M_{\rm Kerr} = 9.59-10.16$. In the SED fitting process we did not account for the data point at Log $\nu \sim 15.4$, probably contaminated by the Lyman $\alpha$ line. Top right panel: SED of the blazar SDSS J013127.34-032100.1 along with a KERRBB model (blue line). The solid blue line has $\nu_{\rm p} \approx 15.21$ and $\nu_{\rm p} L_{\nu_{\rm p}} \approx 47.34$. The relative SS mass is Log$M_{\rm SS} = 9.91$; the KERRBB mass limits, corresponding to $a = -1$ and $a = 0.9999$ are Log$M_{\rm Kerr} = 9.54-10.11$. The solid black line is the continuum extrapolated from \citet{Yietal}. Bottom left panel: SED of the blazar SDSS J074625.87+254902.1 along with a KERRBB model (blue line). The model has $\nu_{\rm p} \approx 15.38$ and $\nu_{\rm p} L_{\nu_{\rm p}} \approx 46.31$. The SS mass is Log$M_{\rm SS} = 9.07$; the KERRBB mass limits, corresponding to $a = -1$ and $a = 0.9999$, are Log$M_{\rm Kerr} = 8.70-9.26$. The solid black line is the continuum extrapolated from the SDSS catalog.  Bottom right panel: SED of the blazar SDSS J161341.06+341247.8 along with a KERRBB model (blue line). The model has $\nu_{\rm p} \approx 15.22$ and $\nu_{\rm p} L_{\nu_{\rm p}} \approx 46.46$. The SS mass is Log$M_{\rm SS} = 9.45$; the KERRBB mass limits, corresponding to $a = -1$ and $a = 0.9999$, are Log$M_{\rm Kerr} = 9.08-9.65$. The solid black line is the continuum extrapolated from the SDSS catalog. In all panels, the orange vertical line highlights the Lyman $\alpha$ line position; we did not consider the frequencies higher than this line for our fit. The IR data (Log $\nu < 14.5$) are probably related to the emission of a dusty torus and/or a nonthermal emission (i.e., synchrotron) and we did not account for them for the fit.
\label{fig:fit}} 
\end{figure*}

\begin{figure*} 
\centering
\includegraphics[width=0.47\textwidth]{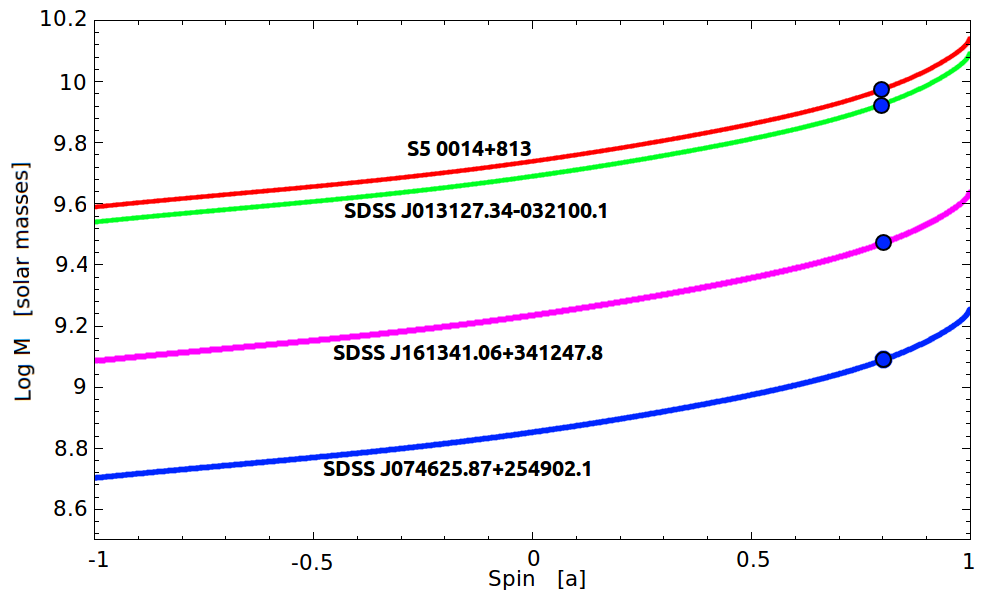} \qquad \includegraphics[width=0.47\textwidth]{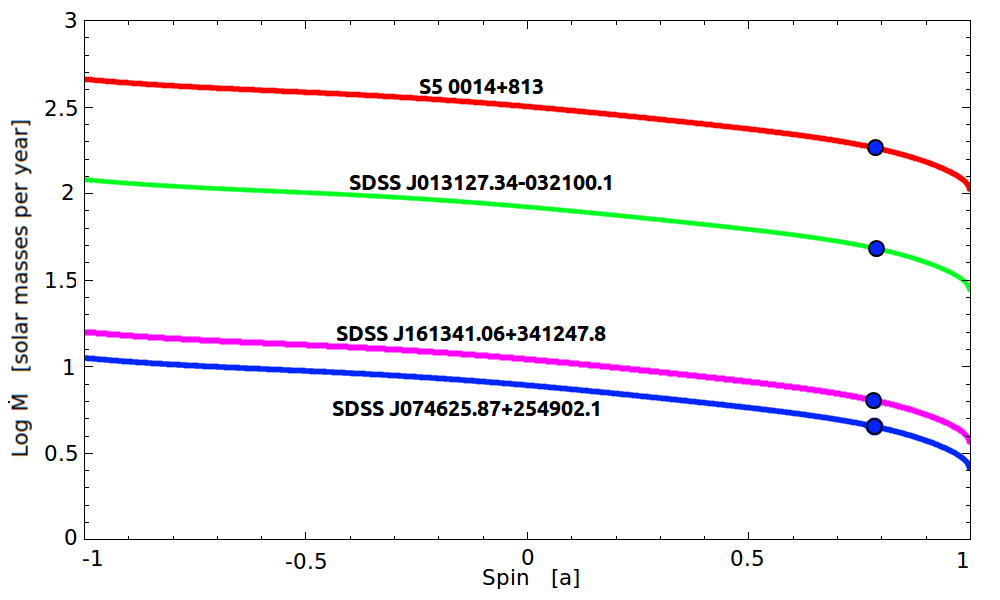}  
\caption{
KERRBB fit solutions of the four blazars. The SS solutions for the same spectrum peak position (blue dots) correspond to the KERRBB solutions with $a \sim 0.8$ (see Table \ref{fitdata} for the peak position and other parameters). KERRBB black hole mass (accretion rate) is larger (smaller) than the SS value for $a \gsim 0.8$.
}
\label{fig:solutions}  
\end{figure*}

\begin{figure} 
\centering
\includegraphics[width=0.47\textwidth]{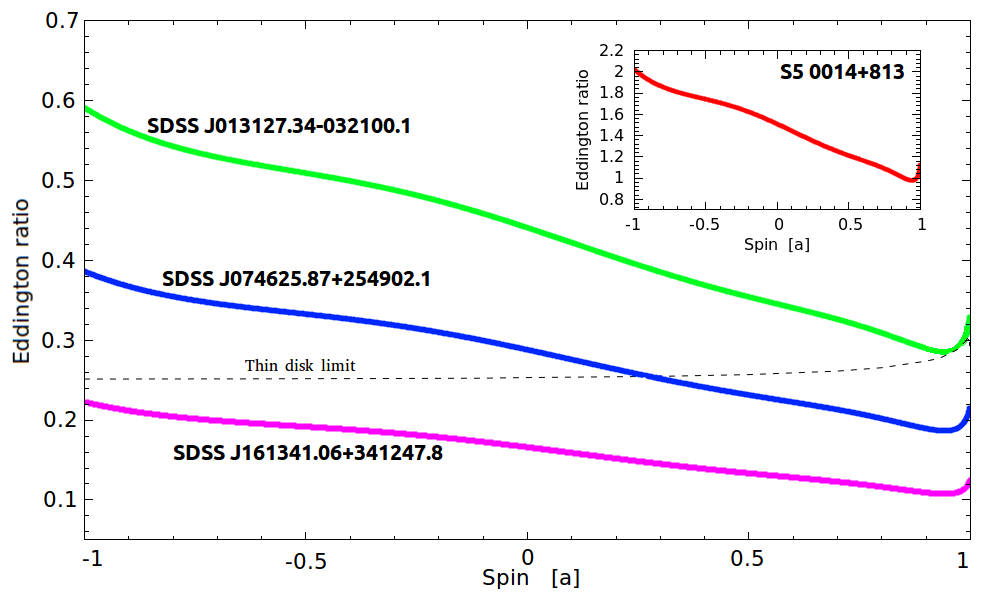}  
\caption{KERRBB Eddington ratio of the sources as a function of the black hole spin ($\theta = 0^{\circ}$). The dotted line represents the thin disk Eddington ratio limit following LN89. According to this limit, the results could be trustworthy for only two sources because of the thin nature of the disk.
}
\label{fig:eddratio}  
\end{figure}

\begin{table*} 
\centering
\footnotesize
\begin{tabular}{llllllllllll}
\hline
Source name & $\nu_{\rm p}$ & \hspace{5mm} $\nu_{\rm p} L_{\nu_{\rm p}}$ & \hspace{5mm} $M_{\rm 0.9}$ & \hspace{5mm} $M_{\rm 0.9982}$ & \hspace{5mm} $\dot{M}_{\rm 0.9}$ & \hspace{5mm} $\dot{M}_{0.9982}$ & \hspace{5mm} $\ell_{\rm Edd, 0.9}$ & \hspace{5mm} $\ell_{\rm Edd, 0.9982}$\\ 
$[1]$ & $[2]$ & \hspace{5mm} $[3]$ & \hspace{5mm} $[4]$ & \hspace{5mm} $[5]$ & \hspace{5mm} $[6]$ & \hspace{5mm} $[7]$ & \hspace{5mm} $[8]$ & \hspace{5mm} $[9]$\\
\hline 
S5 0014+813 & 15.33	& \hspace{5mm} 47.92 & \hspace{5mm} 10.04 & \hspace{5mm} 10.14 & \hspace{5mm} 153.3 & \hspace{5mm} 106.4 & \hspace{5mm} 0.99 & \hspace{5mm} 1.12 \\

SDSS J013127.34-032100.1 & 15.21 & \hspace{5mm} 47.34 & \hspace{5mm} 9.99 & \hspace{5mm} 10.09 & \hspace{5mm} 40 & \hspace{5mm} 27.8 & \hspace{5mm} 0.29 & \hspace{5mm} 0.33 \\

SDSS J074625.87+254902.1 & 15.38 & \hspace{5mm} 46.31 & \hspace{5mm} 9.15 & \hspace{5mm} 9.25 & \hspace{5mm} 3.8 & \hspace{5mm} 2.6 & \hspace{5mm} 0.19 & \hspace{5mm} 0.21 \\

SDSS J161341.06+341247.8 & 15.22 & \hspace{5mm} 46.46 & \hspace{5mm} 9.53 & \hspace{5mm} 9.64 & \hspace{5mm} 5.3 &  \hspace{5mm} 3.7 & \hspace{5mm} 0.11 & \hspace{5mm} 0.12 \\
\hline \\
\end{tabular}
\caption{Results from the fit of four blazars. [1] Name of the source; [2] Log of the spectrum peak frequency in units of Hz; [3] Log of the spectrum peak luminosity in units of erg s$^{-1}$; [4] [5] Log of the KERRBB BH mass related to $a = 0.9$ and $a = 0.9982$ in units of solar masses; [6] [7] Log of the KERRBB accretion rate related to $a = 0.9$ and $a = 0.9982$ in units of $M_{\odot}$yr$^{-1}$; [8] [9] the Eddington ratio related to $a = 0.9$ and $a = 0.9982$. Following LN89, the Eddington ratio is above or near the limit of the thin disk reliability for only two sources ($\ell_{\rm Edd} < 0.3$ $\to$ Fig. \ref{fig:eddratio}).}
\label{fitdata}
\end{table*}

\section{Application to blazars} \label{appl}

To apply our method, we choose a class of sources for which we know the inclination and for which we believe that their black holes
are rapidly spinning. As explained above, this allows us to constrain their masses and accretion rates.

We selected the blazar S5 0014+813 ($z = 3.366$), already studied by \citet{Kuhretal1}, \citet{Kuhretal2}, \citet{Ghi2009S5} and \citet{Sbaretal}; the blazar SDSS J013127.34-032100.1 ($z=5.18$), already studied by \citet{Yietal} and \citet{Ghi2015}; and the blazars SDSS J074625.87+254902.1 and SDSS J161341.06+341247.8, both from the Fermi-LAT and SDSS catalogs (\citealt{Shenetal11}; \citealt{Shawetal}). Their high redshifts shift the accretion disk peak in an observable frequency range, namely in the optical band.

In the SED fitting process, we assume a disk inclination angle of $\theta = 0^{\circ}$ appropriate for blazars: if a larger angle is used (but still $< 5^{\circ}$), the results do not change significantly. We did not consider the frequencies higher than the Lyman $\alpha$ line (Log $\nu = 15.4$) because a prominent absorption feature is usually present due to intervening clouds absorbing hydrogen Lyman alpha photons at wavelengths $<1216$ \AA. Also, we did not account for IR data points (Log $\nu < 14.5$) probably related to the emission of a dusty torus and/or a nonthermal emission (i.e. synchrotron).

Figure \ref{fig:fit} shows the SED-fit of the four blazars along with the Lyman $\alpha$ line position (orange line). The SED-fitting solutions for the black hole mass $M_{\rm Kerr}$ and accretion rate $\dot{M}_{\rm Kerr}$ are shown in Fig. \ref{fig:solutions}. The relative SS solutions are shown with blue dots. Figure \ref{fig:eddratio} shows the Eddington ratio of the sources as a function of the black hole spin.

\subsection{S5 0014+813} 

The SS mass and accretion rate are Log $M_{\rm SS} = 9.96$ and $\dot{M}_{\rm SS} = 182.9\ M_{\odot} \text{yr}^{-1}$, with the spectrum peak in the same position: \citet{Sbaretal} gave a confidence range of Log $M_{BH} = 9.87 - 10$, using a SS model. Assuming a black hole spin $a \lsim 0.8$ ($a \gsim 0.8$), the KERRBB mass (accretion rate) is smaller (larger) than the SS value. If the black hole has a spin $a = 0.9$, the mass will be Log $M_{\rm 0.9} = 10.04$; if is maximally spinning ($a = 0.9982$), the mass will be Log $M_{\rm 0.9982} = 10.14$. The accretion rate $\dot{M}_{\rm 0.9} = 153.3\ M_{\odot} \text{yr}^{-1}$ and $\dot{M}_{\rm 0.9982} = 106.4\ M_{\odot} \text{yr}^{-1}$. In the SED-fitting process we did not account for the photometric point at Log $\nu \sim 15.4$ (Fig. \ref{fig:fit}, left panel), probably contaminated by the Lyman $\alpha$ line. Through the SED fitting process, we found that for all spin values, the Eddington ratio of the source is $\ell_{\rm Edd} \gsim 0.95$ (Fig. \ref{fig:eddratio}), in contrast with LN89 and the Eddington ratio limit for thin disks. For this reason, a different model must be used given the high accretion rates (e.g. slim disk).

\subsection{SDSS J013127.34-032100.1} 

The SS mass and accretion rate are Log $M_{\rm SS} = 9.91$ and $\dot{M}_{\rm SS} = 47.8\ M_{\odot} \text{yr}^{-1}$, with the spectrum peak in the same position: \citet{Ghi2015} found a range of Log $M_{BH} = 9.95 - 10.11$ using a SS model. If the black hole has a spin $a = 0.9$, the mass will be Log $M_{\rm 0.9} = 9.99$; if maximally spinning ($a = 0.9982$), the mass will be Log $M_{\rm 0.9982} = 10.09$. The accretion rate $\dot{M}_{\rm 0.9} = 40\ M_{\odot} \text{yr}^{-1}$ and Log $\dot{M}_{\rm 0.9982} = 27.8\ M_{\odot} \text{yr}^{-1}$. \citet{Yietal} estimated the virial mass of the blazar using MgII line: they found $2.7^{+0.5}_{-0.4} \times 10^{9} M_{\odot}$ (Log $M_{\rm vir} = 9.43^{+0.08}_{-0.07}$) with systematic uncertainty of $\sim 0.4-0.5$ dex. That estimate is smaller than the value found in this work, but compatible if one considers the large systematic uncertainty on the virial mass. Regarding the Eddington ratio, it is always $\ell_{\rm Edd} < 0.59$ and, following LN89, it is easy to see in Fig. \ref{fig:eddratio} that this source is close the limit of the thin disk reliability; as for S5 0014+813, a further check with a different model is necessary.

\subsection{SDSS J074625.87+254902.1}
The SS mass and accretion rate are Log $M_{\rm SS} = 9.07$ and $\dot{M}_{\rm SS} = 4.5\ M_{\odot} \text{yr}^{-1}$, with the spectrum peak in the same position. If the black hole has a spin $a = 0.9$, the mass will be Log $M_{\rm 0.9} = 9.15$; if maximally spinning ($a = 0.9982$), the mass will be Log $M_{\rm 0.9982} = 9.25$. The accretion rate $\dot{M}_{\rm 0.9} = 3.8\ M_{\odot} \text{yr}^{-1}$ and Log $\dot{M}_{\rm 0.9982} = 2.6\ M_{\odot} \text{yr}^{-1}$. From the \citet{Shenetal11} catalog, the virial mass of the black hole is found from the CIV line: $M_{\rm BH} = 9.59 \pm 0.16$, with systematic uncertainty of $\sim 0.4-0.5$ dex. This estimate is larger than our result, but compatible if one considers the large systematic uncertainty on the virial mass. The Eddington ratio is always $\ell_{\rm Edd} < 0.38$; following the LN89 Eddington ratio limit, one can find that for $a < 0.25$ the thin disk approximation is no longer valid (Fig. \ref{fig:eddratio}).

\subsection{SDSS J161341.06+341247.8}
The SS mass and accretion rate are Log $M_{\rm SS} = 9.45$ and $\dot{M}_{\rm SS} = 6.3\ M_{\odot} \text{yr}^{-1}$, with the spectrum peak in the same position. If the black hole has a spin $a = 0.9$, the mass will be Log $M_{\rm 0.9} = 9.53$; if maximally spinning ($a = 0.9982$), the mass will be Log $M_{\rm 0.9982} = 9.64$. The accretion rate will be $\dot{M}_{\rm 0.9} = 5.3\ M_{\odot} \text{yr}^{-1}$ and Log $\dot{M}_{\rm 0.9982} = 3.7\ M_{\odot} \text{yr}^{-1}$. From the \citet{Shenetal11} catalog, the virial mass of the black hole is found from the MgII line: $M_{\rm BH} = 9.69 \pm 0.02$  with systematic uncertainty of $\sim 0.4-0.5$ dex. This estimate is slightly larger than our result. The Eddington ratio is always $\ell_{\rm Edd} < 0.23$, in agreement with LN89.


\section{Summary and conclusions} \label{sec-concl}

In this work, we studied the radiation and emission pattern from an accretion disk around a spinning black hole. We built an analytic approximation of the numerical model KERRBB \citep{Lietal}, developed for X--ray binaries and accounting for all the relativistic effects. With this approach we managed to extend its use to supermassive black holes. Then we applied our analytic method to four well-known high-redshift blazars in order to derive a new estimate of their black hole mass and accretion rate.
We hope that this method will allow an easy use of a relativistic and rather complete accretion disk model.

\subsection{Accretion disk emission pattern}

In \S \ref{sec:empat}, we studied how the observed luminosity $L^{\rm obs}_{\rm d}$ of the accretion disk depends on the viewing angle and the BH spin. We obtained a phenomenological function $f(\theta,a)$ in Eq. \ref{eq:kerrfinale} that closely approximates the variation of the observed luminosity at different $\theta$ and $a$ values. The availability of an analytic expression for the bolometric luminosity related to a spinning black hole can be extremely useful in the analysis of the disk emission features. The pattern can be visualized in Fig. \ref{fig:grid1} which shows the following:
\begin{itemize}
	\item A larger spin implies a larger bolometric luminosity. The increase of luminosity and the strength of the relativistic effects on the pattern are more pronounced for $a \gsim 0.8$;	
	\item At a fixed viewing angle, there is a KERRBB model equivalent to a SS one with the same parameters (mass and accretion rate). At $\theta = 0^{\circ}$, the equivalent KERRBB has spin $a \sim 0.8$. In fact, in order to emit the same luminosity, the KERRBB emission must have a larger efficiency, hence a larger BH spin value. In this way, the two models are equivalent around $\theta = 0^{\circ}$ but, at larger viewing angles, the KERRBB model is brighter because of the strong relativistic effects;
	\item Relativistic effects modifies the pattern at different viewing angles: the simple $\cos\theta$--law (followed by the SS model) is no longer valid, mostly due to light-bending. Hence the maximum observed luminosity is no longer at $\theta = 0^{\circ}$ but at larger viewing angles for larger spin values (see Table \ref{tab:angmax});
	\item The relativistic KERRBB disk emission pattern could be a useful tool for studying the disk's surrounding environment (Broad Line Region, dusty torus) which absorbs a fraction of the disk luminosity. In this way it is possible to constrain some of the AGN properties (Campitiello et al. in preparation).
\end{itemize}

\subsection{Scale relations and family of solutions}

Both the SS and KERRBB models rely upon the assumption of a geometrically thin and optically thick disk. Therefore, the KERRBB disk emission scales with black hole mass and accretion rate as in the SS case (\citealt{Caldero}). 
This is used to scale the KERRBB spectrum from a stellar to a supermassive black hole. Using KERRBB results, we derived Eqs. \ref{eq:nupeak2} and \ref{eq:nulnupeak2} to express peak frequency and luminosity of the spectrum as a function of $a$, $M$, $\dot{M}$ and $\theta$. As illustrated in Fig. \ref{fig:figura5b}, for any given viewing angle, the same spectrum can be reproduced by a family of solutions with different values of the parameters. Since the spin values are limited in the range $[-1,0.9999]$, we expect to find limited ranges also for $M$ and $\dot{M}$. 

In general, we have four free parameters and two observables ($\nu_{\rm p}$ and $\nu_{\rm p} L_{\nu_{\rm p}}$). In specific cases, we have some reliable indications about the viewing angle. Therefore, if we independently know one of the three remaining parameters, we can estimate the other two.

\subsection{Application to data: mass and spin estimate}

In \S \ref{appl} we used KERRBB to fit four blazars (S5 0014+813, SDSS J013127.34--032100.1, SDSS J074625.87+254902.1, SDSS J161341.06+341247.8). Their blazar nature guarantees a very small viewing angle. Furthermore, we can confidently assume a large value of the spin of their black holes. Hence, we estimated the SMBH mass considering $a = 0.9$ and $a = 0.9982$. Figure \ref{fig:fit} shows the SED of the two blazars and the model, while Fig. \ref{fig:solutions} shows how $M$ and $\dot{M}$ change as a function of the spin for the entire spin range listed in Table \ref{fitdata}. We obtained Log $M_{\rm BH} \sim 10.05$ and $10.14$ (S5 0014+813, for $a=0.9$ and $0.9982$ respectively), Log $M_{\rm BH} \sim 9.98$ and $10.09$ (for SDSS J0131--0321), Log $M_{\rm BH} \sim 9.15$ and $9.25$ (for SDSS J0746+2549), Log $M_{\rm BH} \sim 9.53$ and $9.64$ (for SDSS J1613+3412). 

These sources have different Eddington ratios for $a \geq 0.9$, from $\sim 1$ to $\sim 0.11$ (Table \ref{fitdata}): the KERRBB solutions related to S5 0014+813, SDSS J013127.34-032100.1 and SDSS J074625.87+254902.1 (only for $a<0.25$) are in contrast with the assumption of LN89 regarding the thin nature of the disk. This suggests that the disk might be slim or thick and it is necessary to check these sources with different models that account for the modification due to the vertical structure and the high accretion rates. Instead the results related to the source SDSS J161341.06+341247.8 are more trustworthy. We conclude that, although the sources are near or below the Eddington limit, the KERRBB model can be used to obtain a good fit of the data. Also, even if this model is incomplete and inappropriate for extreme luminous sources (and/or for certain spin values), it can be used as a means to understand the nature of the disk; in other words, if a source has $\ell_{\rm Edd} > 0.3$ (calculated with KERRBB), this suggests that the thin disk approximation is not valid and one should use a slim/thick disk model (e.g., SLIMBB, Polish doughnuts; \citealt{SadwAbra}; \citealt{Straubetal}; \citealt{Wieletal}).

In principle, we could use our method to find the BH spin once a reliable estimate of the BH mass is available. Currently, the virial BH masses have an average uncertainty of $0.4$ -- $0.5$ dex, which leads only to a poor constraint on $a$. However, when more precise BH mass estimates are available, the method will provide robust spin estimates and thus new insights into BH physics.




\medskip
\let\itshape\upshape

\begingroup
\let\clearpage\relax


\appendix

\section{Shifting equations}

Equations \eqref{eq:shift1} describe the relations that allow us to rescale the initial spectrum with mass $M_{\rm in}$ and $\dot{M}_{\rm in}$ to a new spectrum with final $M_{\rm fin}$ and $\dot{M}_{\rm fin}$, with a fixed black hole spin. If, in the transformation, the Eddington ratio $\ell = L / L_{\rm Edd}$ remains the same, the equations become functions only of the black hole mass. In fact, we can write
	\[
		L_{\rm in} = \eta \dot{M}_{\rm in} c^2 \qquad \qquad L_{\rm fin} = \eta \dot{M}_{\rm fin} c^2
	\]
	\[
		L_{\rm Edd, in} = C M_{\rm in} \qquad \qquad L_{\rm Edd, fin} = C M_{\rm fin}
	\]

\noindent where $C = 6.32 \cdot 10^4$ erg s$^{-1}$ g$^{-1}$. With $\ell_{\rm Edd} = \eta \dot{M} c^2 / CM$, if we assume the same Eddington ratio $\ell_{\rm in} = \ell_{\rm fin}$, the relation between $\dot{M}$ and $M$ can be found and it leads to
	\begin{equation} \label{eq:shift1apnew}
		\frac{\nu_{\rm p, fin}}{\nu_{\rm p, in}} = \Bigl[ \frac{M_{\rm in}}{M_{\rm fin}} \Bigl]^{1/4} \qquad \qquad \qquad
		\frac{\nu L_{\nu_{\rm p, fin}}}{\nu L_{\nu_{\rm p, in}}} = \frac{M_{\rm fin}}{M_{\rm in}}
	\end{equation}


\section{KERRBB Observed luminosity}

\begin{figure}
\centering
\hskip -0.2 cm
\includegraphics[width=0.5\textwidth]{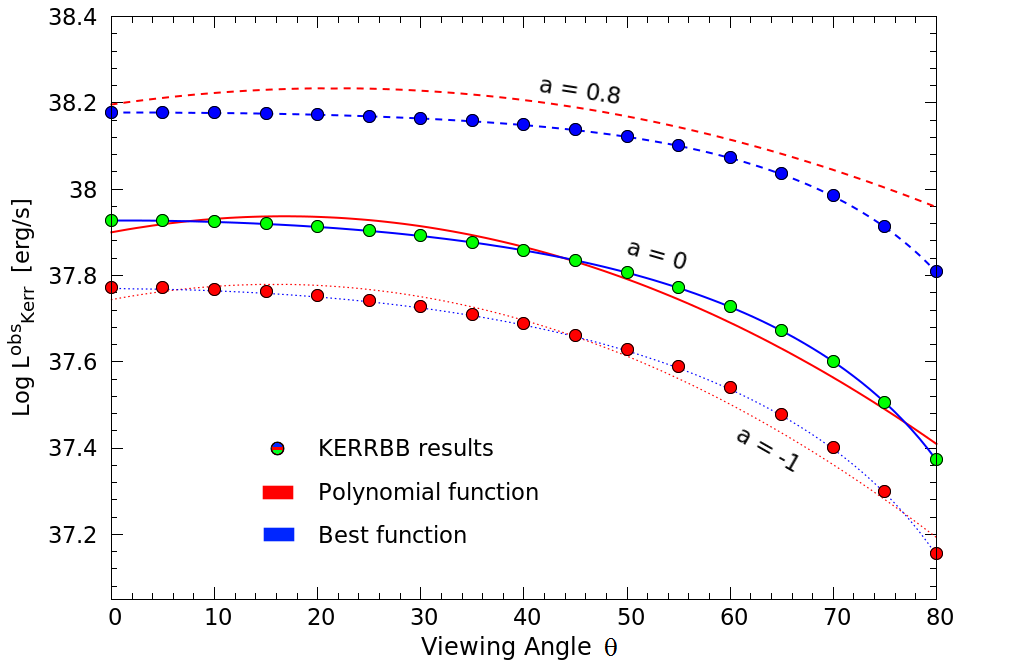}
\caption{KERRBB disk bolometric luminosity as a function of the viewing angle of the disk, fitted by the best function [Eq. \eqref{eq:kerrfinaleapp}, blue lines] and the polynomial function [Eq. \eqref{LBOLPOLY}, red lines] for different spin values. Dots represent the KERRBB results with $\dot{M} = 10^{18}$ g/s. By increasing the spin, the differences between the two functions grow.}
\label{fig:dif}
\end{figure}

In \S \ref{subsec:bollum}, we showed that an analytic expression that approximates the observed frequency integrated luminosity of an accretion disk around a rotating black hole can be written as
	\begin{equation} \label{eq:kerrfinaleapp}
		L^{\rm obs}_{\rm Kerr} (a, \theta) = A \cos \theta [1 - (\sin \theta)^C]^B [1 - E (\sin \theta)^F]^D \ L_{\rm d},
	\end{equation}

\noindent where $L_{\rm d} = \eta(a) \dot{M} c^2$ is the total disk luminosity. All the parameters $A$, $B$, $C$, $D$, $E$, $F$ are functions only of the black hole spin $a$, like the radiative efficiency $\eta(a)$. In order to find the expression for the observed luminosity, for a given value of $a$, a set of viewing angles $\theta$ has been chosen (from $0^{\circ}$ to $85^{\circ}$), and the integrals over frequency of the KERRBB spectra have been computed. We fitted the KERRBB spectra and found the empirical expression \eqref{eq:kerrfinaleapp}. Figure \ref{fig:intang} shows the bolometric luminosities of the KERRBB spectra as a function of the viewing angle, for spin $a = -1$, $-0.6$, $0.6$ and $0.998$ (red dots), and the fitting function \eqref{eq:kerrfinaleapp} (blue line). We note the different behaviors: in the cases $a = -1$, $-0.6$ and $0.6$, the bolometric luminosity between the cases $0^{\circ}$ and $85^{\circ}$ decreases by a factor of $\sim 7$, $6.6$ and $4.4$, respectively\footnote{In the case with no relativistic effects (i.e., SS), there would have been a factor of $\sim 11.5$ between the cases $0^{\circ}$ and $85^{\circ}$.}. In the case with $a = 0.998$, the strong relativistic effects cause the luminosity to reach a maximum value at $\sim 64^{\circ}$ (see Table~\ref{tab:angmax}), and then make it drop at larger viewing angles. The cases with $0^{\circ}$ and $85^{\circ}$ have almost the same luminosity (see also Figure~\ref{fig:spinang}), contrary to the small-spin cases. We note how the residuals are always on the order of $\sim 0.01 \%$, hence function~\eqref{eq:kerrfinaleapp} represents a very good approximation.

Figure \ref{fig:intang2} shows the parameters $A$, $B$, $C$, $D$, $E$ and $F$ for different values of the spin in order to find their dependence on it. As shown in \S \ref{subsec:bollum}, the common fitting equation (blue line) can be written as
	\begin{eqnarray} \label{eq:kerrparaap}
		\mathcal{F}(a) &=& \alpha + \beta x_1 + \gamma x_1^2 + \delta x_1^3 + \epsilon x_1^4 + \iota x_1^5 + \kappa x_1^6
		\nonumber \\
		x_1 &=& \log(1-a)
	\end{eqnarray}

The values of $\alpha$, $\beta$, $\gamma$, $\delta$, $\epsilon$, $\iota$, $\kappa$ are listed Table~\ref{tab:parsping}. we note how the residuals are always on the order of 1$\%$ (or less), except for $F$ (on the order of $\sim 10\%$). These large uncertainties could be reduced using more than six parameters for the fit, but we found it unnecessary for the aim of this work.

\begin{figure*} 
\centering
\includegraphics[width=0.9\textwidth]{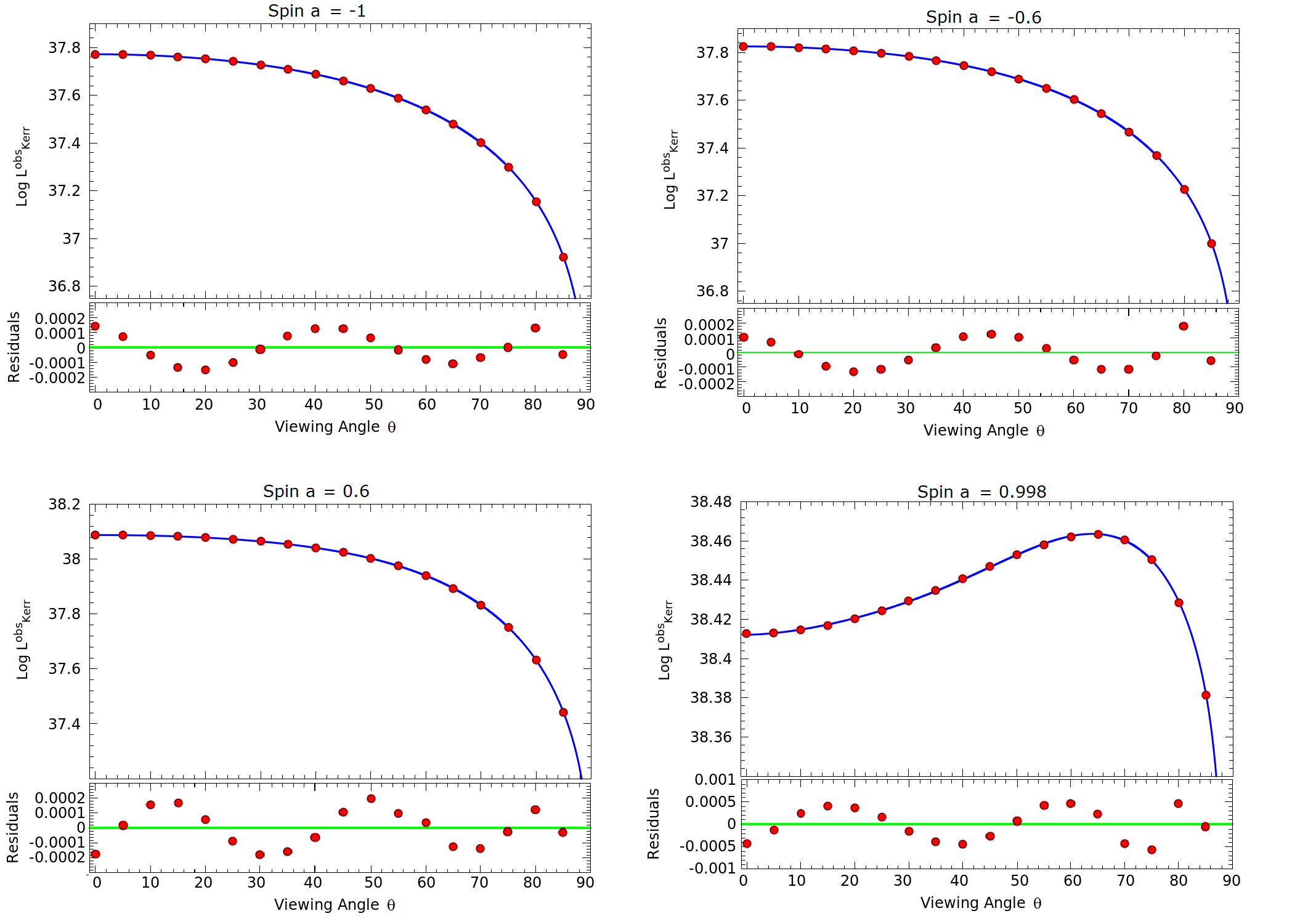}
\caption{KERRBB disk bolometric luminosity (in units of erg/s) as a function of the viewing angle of the disk, in the cases with $a = -1$, $a = -0.6$, $a = 0.6$ and $a = 0.998$, with $\dot{M} = 10^{18}$ g/s. The fitting function (blue line) has the general form of~\eqref{eq:kerrfinale} with different values for the parameters, in the different cases. We note the different behaviors: in the cases $a = -1$, $-0.6$ and $0.6$, the bolometric luminosity, between the cases $0^{\circ}$ and $85^{\circ}$, decreases by a factor of $\sim 7$, $6.6$ and $4.4$, respectively. In the case with $a = 0.998$, the luminosity reaches a maximum value at $\sim 64^{\circ}$ (see Table~\ref{tab:angmax}) then drops at larger viewing angles.      \label{fig:intang}}
\end{figure*}

\begin{figure*} 
\centering
\includegraphics[width=1.04\textwidth]{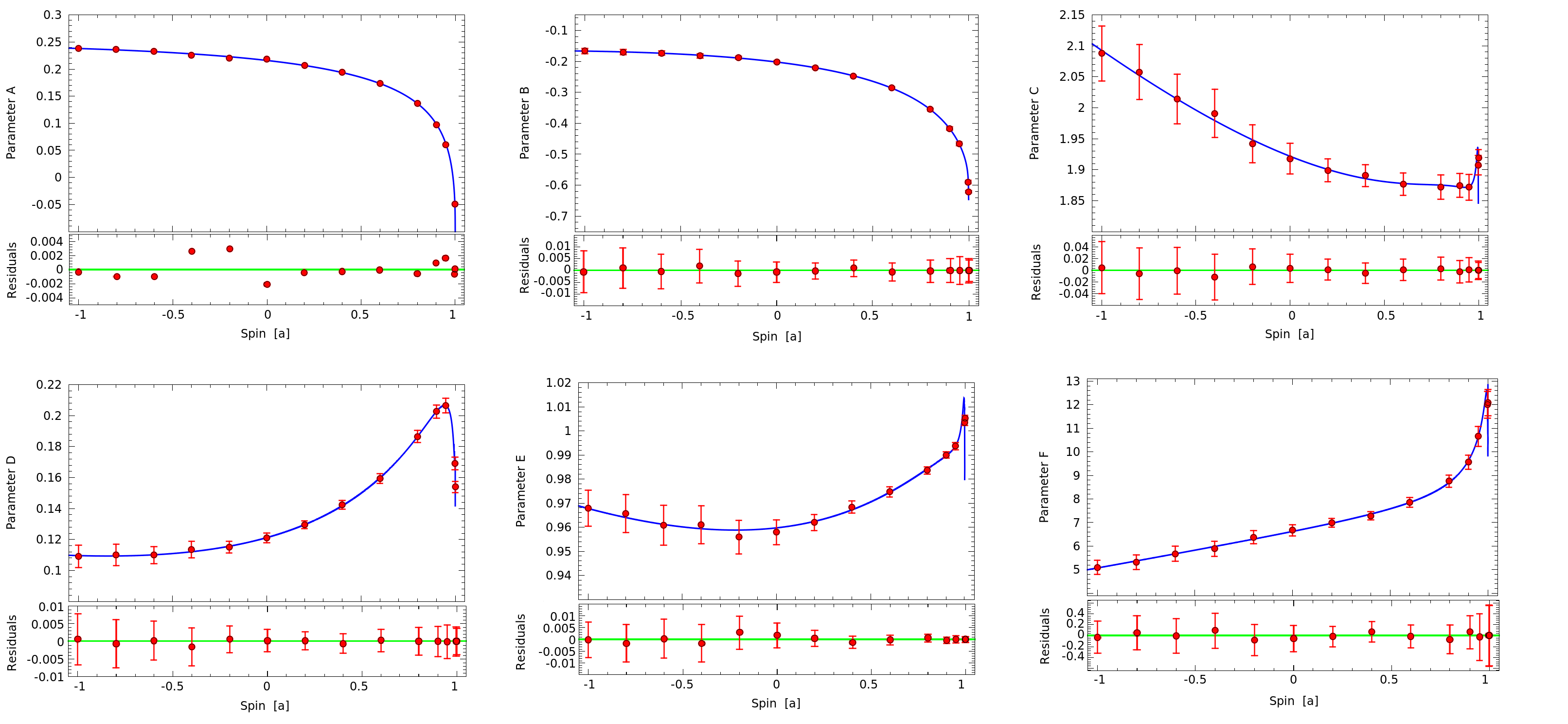}
\caption{Parameters $A$, $B$, $C$, $D$, $E$ and $F$ of Eq. \eqref{eq:kerrfinale} as a function of the black hole spin $a$. The fitting function (blue line) has the general form of Eq. \eqref{eq:kerrpara}. We note how the residuals are always on the order of 1$\%$ (or less), except for $F$ (on the order of $\sim 10\%$). These relatively large residuals for $F$ could be reduced using more than six parameters for the fit, but we found it unnecessary for the aim of this work. Red bars are associated with the uncertainties on the parameter values from the fit of the bolometric luminosity at different viewing angles (Fig. \ref{fig:intang}.)    \label{fig:intang2}}
\end{figure*}

\begin{table*}
\centering
\resizebox{1\textwidth}{!}{
\begin{tabular}{lllllllllll}
\hline 
\textbf{Par of Eq. \eqref{eq:kerrparaap}} & \textbf{$\alpha$} & \textbf{$\beta$} & \textbf{$\gamma$} & \textbf{$\delta$} & \textbf{$\epsilon$} & \textbf{$\iota$} & \textbf{$\kappa$} \\
\hline 
Log A & $0.21595 \pm 0.00065$ & $0.09107 \pm 0.00237$ & $-0.05037 \pm 0.00657$ & $-0.02739 \pm 0.00402$ & $-0.00361 \pm 0.00063$ & 0  & 0\\
B & $-0.20229 \pm 0.00039$ & $0.17538 \pm 0.00169$ & $-0.14390 \pm 0.00657$ & $-0.14534 \pm 0.00858$ & $-0.04544 \pm 0.00358$ & $-0.00480 \pm 0.00045$ & 0\\
C & $1.92161 \pm 0.00273$ & $0.27712 \pm 0.01259$ & $0.67368 \pm 0.04704$ & $0.81327 \pm 0.13254$ & $0.48946 \pm 0.12580$ & $0.13591 \pm 0.04499$ & $0.01373 \pm 0.00527$ \\
D & $0.12120 \pm 0.00024$ & $-0.07852 \pm 0.00110$ & $0.08995 \pm 0.00432$ & $0.12746 \pm 0.00569$ & $0.04556 \pm 0.00238$ & $0.00510 \pm 0.00030$ & 0 \\
E & $0.95973 \pm 0.00080$ & $-0.02003 \pm 0.00288$ & $0.09341 \pm 0.01218$ & $0.16775 \pm 0.02872$ & $0.11440 \pm 0.02475$ & $0.03367 \pm 0.00842$ & $0.00351 \pm 0.00096$ \\
F & $6.62190 \pm 0.02892$ & $-3.84845 \pm 0.09932$ & $-3.11662 \pm 0.41485$ & $-3.61394 \pm 0.60905$ & $-1.54083 \pm 0.26548$ & $-0.19834 \pm 0.03454$ & 0 \\
\hline \\
\end{tabular}
}
\caption{Parameter values of Eq. \eqref{eq:kerrpara} [or equivalently \eqref{eq:kerrparaap}]: the latter expression describes the dependence on the black hole spin of the Kerr disk bolometric luminosity \eqref{eq:kerrfinale} [or equivalently \eqref{eq:kerrfinaleapp}]. The uncertainty on each parameter value is from the fit of the bolometric luminosity at different viewing angles.       \label{tab:parsping}}
\end{table*}

\begin{table}
\centering
\begin{tabular}{lllllllllll}
\hline 
$i$ & \hspace{6mm} $\alpha_i$ & \hspace{6mm} $\beta_i$ & \hspace{6mm} $\gamma_i$ \\
\hline
0	& \hspace{6mm} -1.054	& \hspace{6mm} 0.00443	& \hspace{6mm} -0.000132	\\
1	& \hspace{6mm} 0.159	    & \hspace{6mm} 0.000559	& \hspace{6mm} -3.63e-6 \\
2	& \hspace{6mm} 0.148		& \hspace{6mm} -0.000929	& \hspace{6mm} 4.08e-5 \\
3	& \hspace{6mm} 0.145		& \hspace{6mm} -0.00155	& \hspace{6mm} 5.47e-5 \\
\hline \\
\end{tabular}
\caption{Parameter values of Eq. \eqref{LBOLPOLY}, which represents a less accurate ($\sim 10 \%$) but still good approximation of $L^{\rm obs}_{\rm Kerr}$.       \label{tab:parPOLY}}
\end{table}

\subsection{Polynomial approximation of the observed disk luminosity} \label{POLY}

Once $a$, $\dot{M}$ and $\theta$ are fixed, Eq. \eqref{eq:kerrfinaleapp} needs 37 parameters for the description of the observed frequency integrated disk luminosity $L^{\rm obs}_{\rm Kerr}$ with an accuracy of $\lsim 1 \%$. Another approximation for $L^{\rm obs}_{\rm Kerr}$ can be expressed with the following polynomial function:
	\begin{equation} \label{LBOLPOLY}
		\text{Log} L^{\rm obs}_{\rm Kerr, 2}(a, \theta) = \text{Log} [\dot{M} c^2] + \sum_{i = 0}^{3} a^{i} (\alpha_i + \beta_i \theta + \gamma_i \theta^2)
	\end{equation}

The 12 parameters of this function are given in Table \ref{tab:parPOLY}. The function is valid for $a \in [-1:0.998]$ and $\theta \in [0^{\circ}:80^{\circ}]$. The accuracy of this polynomial approximation is $\sim 10 \%$, less accurate than Eq. \ref{eq:kerrfinaleapp} (differences growing with spin, larger for $\theta \gsim 70^{\circ}$), but still good enough to obtain the bolometric disk luminosity for a wide range of spins and angles (Fig. \ref{fig:dif}). However, because of its accuracy, this approximation does not allow the user to study the emission pattern precisely, as Eq. \ref{eq:kerrfinaleapp} does (see Fig. \ref{fig:grid1}).

\section{Relations between mass, accretion rate and spin}

\begin{figure}
\centering
\hskip -0.2 cm
\vspace{1mm} \includegraphics[width=0.498\textwidth]{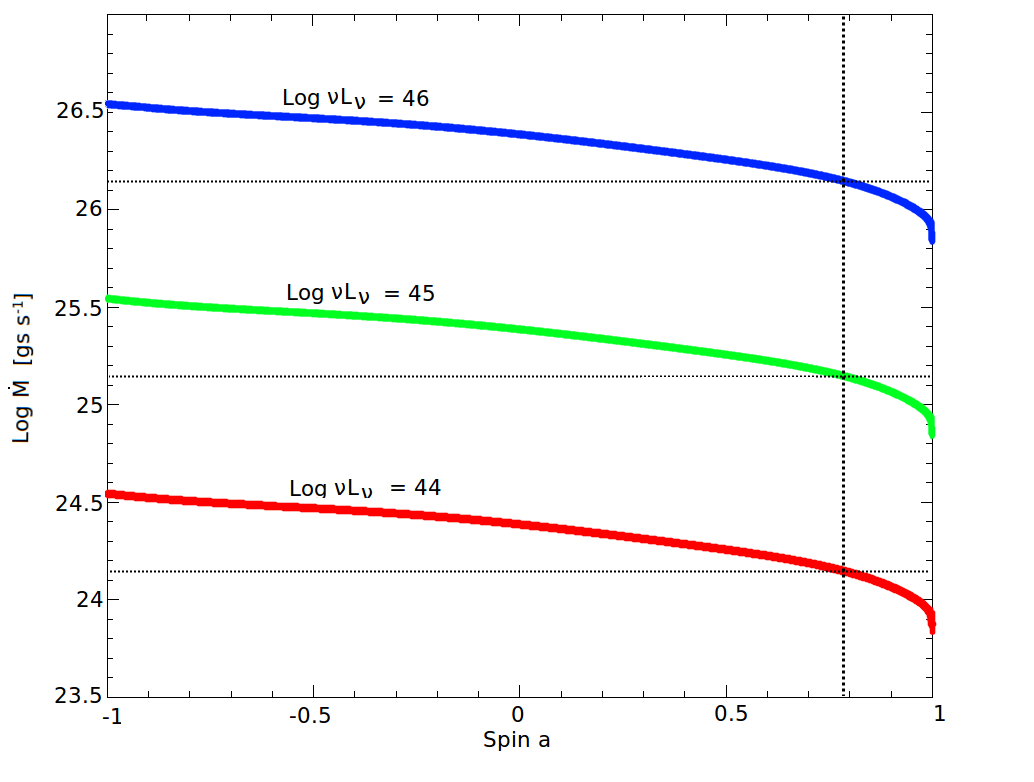} \includegraphics[width=0.497\textwidth]{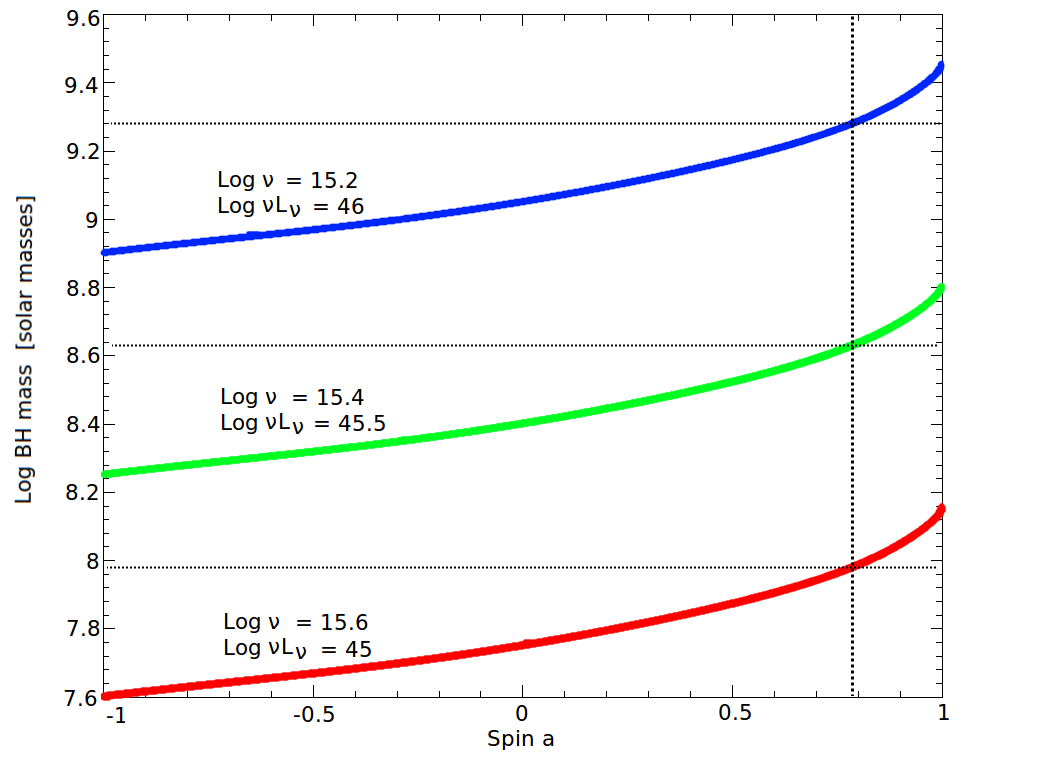}
\caption{Top panel: Relation between the accretion rate $\dot{M}$ and the spin $a$: different lines are related to different values of peak luminosity $\nu_{\rm p} L_{\nu_{\rm p}}$. Bottom panel: Relation between the black hole mass $M$ and the spin $a$: different lines are related to different couples of peak frequency and luminosity $\nu_{\rm p} - \nu_{\rm p} L_{\nu_{\rm p}}$. For every line, the spectrum peak $\nu_{\rm p}$ -- $\nu_{\rm p} L_{\nu_{\rm p}}$ and the viewing angle $\theta=0^{\circ}$ remain fixed. The SS solutions (horizontal dotted lines) with the same peak frequency and luminosity correspond to the KERRBB model with $a \sim 0.8$ (vertical dotted line). Since the spectrum peak moves to higher frequencies and luminosities by increasing the spin value (see Fig. \ref{fig:figura5b}), the mass and the accretion rate must increase and decrease, respectively, in order to keep the spectrum peak in the same position.}
\label{SMMdot}
\end{figure}

As shown before, the corrected equations for the KERRBB peak frequency and luminosity are Eq. \eqref{eq:nupeak2} and Eq. \eqref{eq:nulnupeak2}. By fitting the variation of the spectrum peak position at different spin values and fixed $\theta = 0^{\circ}$, functions $g_1(a, \theta=0^{\circ})$ and $g_2(a, \theta=0^{\circ})$ have been expressed as a functional given by Eq. \eqref{eq:functional2}. As pointed out before, the frequency $\nu_{\rm p}$ and the luminosity $\nu_{\rm p} L_{\nu_{\rm p}}$ are degenerate in mass, accretion rate and spin (if the viewing angle $\theta$ is fixed); the same KERRBB spectrum (i.e., the emission peak in the same position) can be fitted by a family of solutions by changing the black hole mass, accretion rate and spin in the proper way (see Fig. \ref{fig:figura5b}). It is important to note that when the disk viewing angle $\theta$ is fixed, the quantities $M$, $\dot{M}$ and $a$ are connected to each other: knowing the value of one of them fixes the values of the other two. Given a peak position $[\nu_{\rm p}, \nu_{\rm p} L_{\nu_{\rm p}}]$, using Eq. \eqref{eq:nupeak2} and Eq. \eqref{eq:nulnupeak2}, the relations between the black hole mass $M$ the accretion rate $\dot{M}$ and the black hole spin $a$ can be found easily. 
Figure \ref{SMMdot} (left panel) shows the relation between the accretion rate $\dot{M}$ and the black hole spin $a$: it is easy to see that by increasing the spin (i.e., increasing the radiative efficiency $\eta$), the accretion rate must decrease in order to keep the same peak luminosity. The right panel shows the relation between the black hole mass $M$ and the spin $a$: by increasing the spin, the black hole mass must increase as well. Figure \ref{fig:figura5b} is useful to explain this feature: by increasing the spin value, the spectrum peak moves to higher frequencies and luminosities, hence by increasing the mass value and decreasing the accretion rate, the peak moves to lower frequencies and luminosities (i.e. to the initial position). We note that the KERRBB mass (accretion rate) is larger (smaller) than the SS value only for spin values $a \gsim 0.8$.


\label{lastpage}
\endgroup

\end{document}